\documentclass[11pt, a4paper]{article}
\usepackage{jheppub}
\usepackage{epsfig}
\usepackage{amsfonts}
\usepackage{amsmath}
\usepackage{amssymb}
\usepackage{dsfont}
\usepackage{hyperref}
\usepackage{color}
\usepackage{bbm} 

\newcommand{\al}{\alpha}

\newcommand{\si}{\sigma}

\newcommand{\simu}{\sigma^{\mu\nu}}

\newcommand{\Or}{\mathcal O}

\newcommand{\vL}{\ensuremath{\mathcal{L}}}
\newcommand{\vp}{\varphi}
\newcommand{\sq}{^{2}}

\newcommand{\dslash}[1]{#1 \llap{/\kern-0.5pt}}
\newcommand{\Dslash}[1]{#1 \llap{/\kern+1.5pt}}
\newcommand{\DDslash}[1]{#1 \llap{/\kern+2.3pt}}
\newcommand{\dslashh}[1]{#1 \llap{/\kern+1pt}}

\newcommand{\boldtau}{\mbox{\boldmath $\tau$}}

\newcommand{\sw}{s_{0w}}
\newcommand{\cw}{c_{0w}}
\newcommand{\swsq}{s^{2}_{0w}}
\newcommand{\cwsq}{c^{2}_{0w}}

\newcommand{\bea}{\begin{eqnarray}}
\newcommand{\eea}{\end{eqnarray}}
\newcommand{\bma}{\begin{pmatrix}}
\newcommand{\ema}{\end{pmatrix}}
\newcommand{\nn}{\nonumber}
\newcommand{\red}[1]{{\color{red}#1}}

\newcommand{\green}[1]{{\color{green}#1}}

\newcommand\POWHEG{{\tt POWHEG}}
\newcommand\POWHEGBOX{{\tt POWHEG\ BOX}}
\newcommand\POWHEGBOXVTWO{{\tt POWHEG BOX V2}}
\newcommand\PythiaEight{{\tt Pythia8}}

\title{NLO QCD corrections to SM-EFT dilepton and electroweak Higgs boson production, matched to parton shower in POWHEG}
\author[a,b]{Simone Alioli,}
\emailAdd{simone.alioli@unimib.it}
\author[c]{Wouter Dekens,}
\emailAdd{wdekens@lanl.gov}
\author[d]{Michael Girard,}
\emailAdd{michael.girard@berkeley.edu}
\author[c]{Emanuele Mereghetti,}
\emailAdd{emereghetti@lanl.gov}

\affiliation[a]{Universita' degli Studi di Milano-Bicocca,\\ 20126, Milano,
  Italy}
\affiliation[b]{INFN, Sezione di Milano-Bicocca,\\ 20126, Milano, Italy}
\affiliation[c]{Theoretical Division, Los Alamos National Laboratory,\\
Los Alamos, NM 87545, USA}
\affiliation[d]{Ernest Orlando Lawrence Berkeley National Laboratory, University of California, \\
Berkeley, CA 94720, USA}

\abstract{We discuss the Standard Model - Effective Field Theory
  (\mbox{SM-EFT}) contributions to neutral- and charge-current Drell-Yan
  production, associated production of the Higgs and a vector boson, and Higgs
  boson production via vector boson fusion.  We consider all the dimension-six
  \mbox{SM-EFT} operators that contribute to these processes at leading order,
  include next-to-leading order QCD corrections, and interface them with parton
  showering and hadronization in \PythiaEight{} according to the \POWHEG{} method. We discuss
  existing constraints on the coefficients of dimension-six operators and
  identify differential and angular distributions that can differentiate between different effective operators, pointing to specific
  features of Beyond-the-Standard-Model physics.

}

\begin{document}
\maketitle

\section{Introduction}

The Standard Model (SM) of particle physics is an extremely successful theory.  It is in good agreement with the precise measurements at the $Z$ pole \cite{ALEPH:2005ab}, with numerous flavor observables   \cite{Bona:2006ah,Charles:2015gya,Amhis:2016xyh}, the observed CP violation in the kaon and $B$ meson systems \cite{Amhis:2016xyh}, and the non-observation of electric dipole moments \cite{Baker:2006ts,Graner:2016ses,Bishof:2016uqx,Baron:2013eja,Cairncross:2017fip}.
Even the hints for deviations from the SM, observed in the muon anomalous magnetic moment \cite{Bennett:2006fi} and in direct CP violation in kaon decays \cite{Bai:2015nea},
seem to indicate  that the scale of new physics is high, in the 10 to 100 TeV range. 
In this scenario, the manifestations of  new physics at the Large Hadron Collider (LHC) can be described by non-renormalizable  higher-dimensional operators, 
which are invariant under the SM gauge group and  whose effects are suppressed
by powers of the scale $\Lambda$ that characterizes the new physics. 
Extending the SM with these higher-dimensional operators gives rise to an effective field theory (EFT) in which the lowest-dimensional operators  provide the dominant effects as long as $\Lambda$ is well above the electroweak scale.
If, in addition, we assume there are no new light fields and that the Higgs boson observed at the LHC belongs to an $SU(2)_L$ doublet, this EFT is referred to as the SM-EFT \cite{Buchmuller:1985jz,Grzadkowski:2010es,Jenkins:2013wua,Jenkins:2013zja,Alonso:2013hga}.

The higher-dimensional operators start at dimension-five, where there is a single gauge-invariant operator that violates lepton number and is responsible for the neutrino masses and mixings  \cite{Weinberg:1979sa}. 
The gauge-invariant dimension-six operators were cataloged in Refs.\ \cite{Buchmuller:1985jz,Grzadkowski:2010es}. The basis of dimension-seven operators has been constructed in Ref.\ \cite{Lehman:2014jma}, while 
studies of dimension-eight operators are also starting to appear  \cite{Henning:2015alf}. Finally, a subset of lepton-number violating dimension-nine operators has been classified in the context 
of neutrinoless double beta decay  \cite{Babu:2001ex,deGouvea:2007qla,Graesser:2016bpz}.
The SM-EFT provides a very general parameterization of deviations from the SM at colliders, 
as well as for low-energy observables  such as electric dipole moments \cite{deVries:2012ab,Engel:2013lsa}, 
beta decay 
\cite{Cirigliano:2009wk,Cirigliano:2012ab,Cirigliano:2013xha,Gonzalez-Alonso:2016etj,Falkowski:2017pss,Gonzalez-Alonso:2018omy}, and flavor physics \cite{Alonso:2014csa}.
This, combined with the absence of discoveries of new states at the LHC, has led to increased attention to the phenomenology of the SM-EFT in recent years  \cite{Brivio:2017vri}, particularly in the sectors of the Higgs boson \cite{Artoisenet:2013puc,Maltoni:2013sma,Alloul:2013naa,Demartin:2014fia,Mimasu:2015nqa,Degrande:2016dqg,Maltoni:2016yxb,Ferreira:2016jea,Deutschmann:2017qum,Greljo:2017spw,Ellis:2018gqa} and top quark 
\cite{Zhang:2010dr,Degrande:2012gr,Demartin:2015uha,Franzosi:2015osa,Demartin:2016axk,Bylund:2016phk,Cirigliano:2016nyn,Zhang:2017mls,AguilarSaavedra:2018nen,Degrande:2018fog}.

In this paper we study the complete set of dimension-six operators, in the
basis of Ref.\ \cite{Grzadkowski:2010es}, that gives tree-level contributions
to the neutral-current (NC) and charged-current (CC) Drell-Yan processes, to the associated production of the Higgs and a $W$ or $Z$ boson ($WH$ and $ZH$ production),
and to Higgs boson production via vector boson fusion (VBF).
We include next-to-leading order (NLO) QCD corrections to the partonic processes, and interface with the parton shower according to the \POWHEG{} method \cite{Nason:2004rx,Frixione:2007vw,Alioli:2010xd}, extending the original works of Refs.~\cite{Alioli:2008gx,Nason:2009ai,Luisoni:2013kna}.

SM-EFT analyses of NC and CC Drell-Yan processes have mainly considered
contributions at leading order in QCD
\cite{Cirigliano:2012ab,Gonzalez-Alonso:2016etj,Brivio:2017btx,Brivio:2017vri,Falkowski:2017pss,Farina:2016rws}
or included higher-order QCD corrections via a form-factor rescaling~\cite{Greljo:2017vvb,Alioli:2017nzr}. Here
we overcome this limitation by including generic NLO QCD corrections, and interfacing
to the parton showers. This allows us to reliably estimate the theoretical
error on the SM-EFT cross section.  It also provides a tool for the study of
generic observables, more differential than the dilepton invariant or
transverse mass. Some of these observables, such as  the lepton
angular distributions, can be useful to disentangle the contributions from
different effective operators and could therefore provide important clues on the
nature of the new physics, were deviations from the SM to be observed in 
Drell-Yan production.

For Higgs boson production, NLO QCD corrections to the contributions to $WH$, $ZH$, and VBF of a subset of SM-EFT operators have been studied in Refs.\  \cite{Maltoni:2013sma,Demartin:2014fia,Mimasu:2015nqa,Degrande:2016dqg,Maltoni:2016yxb,Greljo:2017spw}. Here we consider a slightly  larger set of operators, with less restrictive flavor assumptions for the couplings to quarks. Also in this case, 
including the effective operators in \POWHEG{} allows for the study of
differential and angular distributions, which could help disentangle the
effects of different dimension-six operators.

The paper is organized as follows. In Section \ref{basis} we introduce the operator basis, and establish our notation for the coefficients of the dimension-six operators we implemented in \POWHEG{}.
We discuss the NC and CC Drell-Yan processes in Section \ref{Wprod}, while  in Sections \ref{HVprod} and \ref{VBF}
we describe, respectively,  the associated production of a Higgs boson and a $W$ or $Z$ boson, and Higgs production via vector boson fusion, before concluding in Section \ref{conclusion}. 
We relegate  several SM-EFT corrections to the Higgs boson decay to Appendix \ref{AppDecay}.

\section{The operator basis}\label{basis}
Before discussing dimension-six operators, we recall a few SM ingredients needed to establish our  conventions.  
The SM Lagrangian is completely determined  by the invariance under the Lorentz group, the gauge group $SU(3)_c \times SU(2)_L \times U(1)_Y$, 
and by the matter content. We consider here the SM in its minimal version, with three families of leptons and quarks, and one scalar doublet. 
The left-handed quarks and leptons transform as doublets under $SU(2)_L$
\begin{equation}
q_L  = \left( \begin{array}{c}
u_L \\
d_L
\end{array}
\right), \qquad \ell_L = \left( \begin{array}{c}
\nu_L \\
e_L
\end{array}
\right),
\end{equation}
while the right-handed quarks, $u_R$ and $d_R$, and charged leptons, $e_R$,
are singlets under $SU(2)_L$. We do not include  sterile right-handed
neutrinos, but their effects on e.g.  $W$ production can be straightforwardly included \cite{Cirigliano:2012ab}.
The scalar field $\varphi$ is a doublet under $SU(2)_L$. In the unitary gauge we have
\begin{equation}
\varphi = \frac{v}{\sqrt{2}} U(x) \left( 
\begin{array}{c}
0 \\
1 +  \frac{h}{v}
\end{array}\right),
\end{equation}
where $v=246$ GeV is the scalar vacuum expectation value ($vev$), $h$ is the physical Higgs field and $U(x)$ is a unitary matrix that encodes the Goldstone bosons. By  $\tilde \vp$ we denote $\tilde \vp = i\tau_2 \vp^*$. 

The gauge interactions are determined by the covariant derivative
\begin{equation}
D_\mu =  \partial_\mu + i  \frac{g}{2} \boldtau \cdot W_\mu + i g^\prime Y B_\mu  + i g_s G^a_\mu t^a  
\end{equation}
where $B_\mu$, $W^I_{\mu}$ and $G^a_{\mu}$ are the $U(1)_Y$, $SU(2)_L$ and $SU(3)_c$ gauge fields, respectively, and $g'$, $g$, and $g_s$ are their gauge couplings. Furthermore, 
$\boldtau/2$ and $t^a$ are the $SU(2)_L$ and $SU(3)_c$ generators, in the representation of the field on which the derivative acts.
In the SM, the gauge couplings $g$ and $g^\prime$ are related to the electric charge and the Weinberg angle by $g s_w = g^\prime c_w = e$, where $e > 0$ is the charge of the positron
and $s_w = \sin\theta_W$, $c_w = \cos\theta_W$. We will shortly discuss how these relations are affected in the presence of dimension-six operators.
The hypercharge assignments under the group  are $1/6$, $2/3$, $-1/3$, $-1/2$, $-1$, and $1/2$ for $q_L$,
$u_R$ , $d_R$ , $\ell_L$ , $e_R$ , and $\vp$, respectively. The SM Lagrangian then consists of the Lorentz- and gauge-invariant terms with dimension $d\leq 4$ that can be constructed from the above fields.

The processes we aim to study, Drell-Yan, $WH$, $ZH$, and VBF, are affected by many dimension-six operators. Following the notation of Ref.\ \cite{Grzadkowski:2010es}, we classify the relevant operators 
according to their content of gauge (denoted by $X$), fermion ($\psi$), and scalar fields ($\varphi$). The operators that contribute  at tree level fall in the following five classes 
\begin{eqnarray}\label{eq:basis1}
\mathcal L =   \mathcal L_{X^2 \varphi^2} + \mathcal L_{\psi^2 X \varphi} + \mathcal L_{\psi^2 \varphi^2 D} + \mathcal L_{\psi^2 \varphi^3} + \mathcal L_{\psi^4}.   
\end{eqnarray}
Here $\mathcal L_{X^2 \varphi^2}$ contains operators with two scalars and two gauge bosons. At the order we are working and for the processes we are considering, the only relevant operators are the ones involving  $SU(2)_L$ and $U(1)_Y$ gauge bosons:
\begin{eqnarray}\label{eq:Xphi2}
\mathcal L_{X^2 \varphi^2} &=& C_{\varphi W}\, \frac{ \varphi^{\dagger} \varphi}{v^2} W^I_{\mu \nu} W^{I\, \mu \nu}  +  C_{\varphi B}\, \frac{ \varphi^{\dagger} \varphi}{v^2} B_{\mu \nu} B^{ \mu \nu}
+ C_{\varphi W B}\, \frac{\varphi^{\dagger} \tau^I \varphi}{v^2} W^I_{\mu \nu} B^{\mu \nu} \nn \\
& + &  C_{\varphi \tilde{W}}\, \frac{\varphi^{\dagger} \varphi}{v^2} \tilde{W}^I_{\mu \nu} W^{I\,\mu \nu} +  C_{\varphi \tilde B}\, \frac{\varphi^{\dagger} \varphi}{v^2} \tilde{B}_{\mu \nu} B^{ \mu \nu}
+ C_{\varphi \tilde W B}\, \frac{\varphi^{\dagger} \tau^I \varphi}{v^2} \tilde{W}^I_{\mu \nu}\, {B}^{\mu \nu} 
\end{eqnarray}
where $W^I_{\mu\nu}$ and $B_{\mu\nu}$ denote the  $SU(2)_L$ and $U(1)_Y$ field strengths,
and $\tilde X_{\mu \nu} = \varepsilon_{\mu \nu \alpha \beta} X^{\alpha
  \beta}/2$. 
The operators in the first line of Eq.\ \eqref{eq:Xphi2} are CP-even, while those on the second line violate CP.
Here, and in what follows, we define the coefficients of dimension-six operators to be dimensionless. Thus, $C_{\varphi X}$ and $C_{\varphi \tilde X}$ scale as $v^2/\Lambda^{2}$, where $\Lambda$ is the new physics scale. 
The remaining operators in the class  $\mathcal L_{X^2
  \varphi^2}$ involve the gluon field strength $G_{\mu \nu}$. While these are
very interesting for Higgs boson production via gluon fusion, they contribute to the
processes we are considering -- all of which are (anti-)quark-initiated -- only at higher order, and we neglect them. 

The classes $\psi^2 X \varphi$ and $\psi^2 \varphi^2 D$ both contain fermion bilinears. The former class consists of dipole operators, of which we focus on the dipole couplings of quarks and leptons to the $SU(2)_L$ and $U(1)_Y$ gauge bosons:   
\begin{eqnarray}\label{eq:dipole}
\mathcal L_{\psi^2 X \varphi} &=&  -\frac{1}{\sqrt{2}}\bar q_L \si^{\mu\nu}(g^\prime\Gamma_{B}^u B_{\mu\nu}+g\Gamma_{W}^u \boldsymbol{\tau} \cdot \boldsymbol{W}_{\mu\nu})
 \frac{\tilde \vp}{v^2 }u_R \nn\\
& &-\frac{1}{\sqrt{2}}\bar q_L \si^{\mu\nu}(g^\prime\Gamma_{B}^d B_{\mu\nu}+g\Gamma_{W}^d \boldsymbol{\tau} \cdot \boldsymbol{W}_{\mu\nu})\frac{\varphi}{v^2}  d_R \nn\\
& &-\frac{1}{\sqrt{2}}\bar \ell_L \si^{\mu\nu}(g^\prime\Gamma_{B}^e B_{\mu\nu}+g\Gamma_{W}^e \boldsymbol{\tau} \cdot \boldsymbol{W}_{\mu\nu})\frac{\varphi}{v^2}  e_R + \mathrm{h.c.}\ . 
\end{eqnarray}
Here $\Gamma_W^{u,d,e}$ and $\Gamma_B^{u,d,e}$ are generally $3\times 3$ matrices in flavor space,  which we will discuss in more detail shortly.
In what follows we trade the couplings to the $B$ gauge field,   $\Gamma_B^{u,d,e}$, for those to the  photon field
\begin{equation}
\Gamma^u_\gamma = \Gamma^u_B  + \Gamma^u_W, \qquad  \Gamma^d_\gamma = \Gamma^d_B  - \Gamma^d_W \quad \textrm{and} \quad  \Gamma^e_\gamma = \Gamma^e_B  - \Gamma^e_W.
\end{equation}

$\mathcal L_{\psi^2 \varphi^2 D}$ contains corrections to the SM couplings of the quarks and leptons to the $Z$ and $W$ bosons, which, in gauge invariant form, appear as
\bea
\vL_{\psi^2 \varphi^2 D} &=&   -  \frac{\varphi^{\dagger} i \overleftrightarrow{D}_{\mu} \varphi}{v^2} \, \left(   \bar q_L   \gamma^{\mu} \, c^{(1)}_{Q\varphi} q_L +   \bar u_R   \gamma^{\mu}\,  c_{U\varphi} u_R 
+  \bar d_R  \gamma^{\mu}\, c_{D\varphi}  d_R\right) \nonumber \\
& & -  \frac{ \varphi^{\dagger}  i \overleftrightarrow{D}^I_{\mu} \varphi}{v^2}\,  \bar q_L \tau^I  \gamma^{\mu} c^{(3)}_{Q\varphi} q_L + 
\left( i \frac{2}{v^2} \tilde{\vp}^{\dagger} D_{\mu} \vp \, \bar{u}_R \gamma^\mu \xi d_R 
+  \mathrm{h.c.} \right) \nn \\
& & -  \frac{\varphi^{\dagger} i \overleftrightarrow{D}_{\mu} \varphi}{v^2} \, \left(   \bar \ell_L   \gamma^{\mu} \, c^{(1)}_{L\varphi} \ell_L +   
  \bar e_R  \gamma^{\mu}\, c_{e\varphi}  e_R\right)  -  \frac{ \varphi^{\dagger}  i \overleftrightarrow{D}^I_{\mu} \varphi}{v^2}\,  \bar \ell_L \tau^I  \gamma^{\mu} c^{(3)}_{L\varphi} \ell_L .
\label{eq:vertex}
\eea
Here $\overleftrightarrow{D}_{\mu}=  D_\mu-\overleftarrow D_\mu$, $\overleftrightarrow{D}^I_{\mu}= \tau^I D_\mu-\overleftarrow D_\mu \tau^I$, and $c_{Q\vp}^{(1)}$, $c_{Q\vp}^{(3)}$, $c_{U\vp}$, $c_{D\vp}$, $c_{L\vp}^{(1)}$, $c_{L\vp}^{(3)}$, $c_{e\vp}$
are hermitian, 3 $\times$ 3 matrices, while $\xi$ is a  generic 3 $\times$ 3 matrix.
The operators $c_{L\vp}^{(1)}$, $c_{L\vp}^{(3)}$, $c_{e\vp}$ couple lepton bilinears to the weak bosons, thus affecting $Z$ and $W$ production at tree level. 
However, these operators  are strongly constrained by LEP measurements of the physics at the $Z$ pole \cite{ALEPH:2005ab}. Furthermore, the $Z$ and $W$ production cross sections induced by these operators have the same shape as 
the SM, making it hard to identify them at collider. For these reasons, we neglect them in what follows.

The remaining operators involving quark bilinears in the dimension-six SM-EFT Lagrangian are corrections to the quark and lepton Yukawa couplings
\begin{equation}\label{eq:yuk}
\mathcal L_{\psi^2 \varphi^3} = - \sqrt{2} \frac{\varphi^{\dagger} \varphi}{v^2}  \left( \bar q_L Y^\prime_u \tilde \varphi u_R + \bar q_L Y^\prime_d \varphi d_R + \bar \ell_L Y^\prime_e \varphi e_R   \right).
\end{equation}
These couplings are mainly constrained by Higgs boson production via quark-antiquark annihilation and by the Higgs
boson branching ratios, but we will see that the quark couplings $Y^\prime_{u,d}$ also give interesting contributions to $WH$
and $ZH$. 

Finally, we  consider the following  semileptonic four-fermion operators,
\begin{eqnarray}\label{eq:fourfermion}
\mathcal L_{\psi^4} &=& -\frac{4 G_F}{\sqrt{2}}  \Bigg\{ 
C^{(1)}_{LQ}\, \bar \ell_L \gamma^\mu \ell_L \, \bar q_L \gamma_\mu q_L + C^{(3)}_{LQ} \, \bar \ell_L \boldtau \gamma^\mu \ell_L \,\cdot \bar q_L  \boldtau \gamma_\mu q_L + C_{eu} \, \bar e_R \gamma^\mu e_R \, \bar u_R \gamma_\mu u_R  \nn \\ & & \qquad \qquad \quad +\
 C_{ed} \, \bar e_R \gamma^\mu e_R \, \bar d_R \gamma_\mu d_R + C_{Lu}\, \bar \ell_L \gamma^\mu \ell_L \, \bar u_R \gamma_\mu u_R +  C_{Ld}\, \bar \ell_L \gamma^\mu \ell_L \, \bar d_R \gamma_\mu d_R  \nn \\ 
& &  \qquad \qquad \quad +\ C_{Qe}  \, \bar e_R \gamma^\mu e_R \, \bar q_L \gamma_\mu q_L   \Bigg\}  \\
& &-\frac{4 G_F}{\sqrt{2}} \Bigg\{  C_{LedQ}\, \bar \ell^i_L e_R\, \bar d_R q_L^i + C^{(1)}_{LeQu}\, \varepsilon^{ij} \bar \ell^i_L e_R\, \bar q_L^j u_R +  C^{(3)}_{LeQu}\, \varepsilon^{ij} \bar \ell^i_L \sigma^{\mu \nu} e_R\, \bar q_L^j \sigma_{\mu \nu} u_R \, + \textrm{h.c.}
  \Bigg\}. \nonumber
\end{eqnarray}
Of these operators, only a few affect charged currents, introducing new Lorentz structures, such as scalar-scalar and tensor-tensor interactions. All of the above operators modify neutral currents and the couplings are, in general, four-indices tensors. 

Having introduced the dimension-six operators of interest, we will first discuss how these new interactions affect the SM couplings. The assumptions on the flavor structure of the introduced couplings are discussed in Section \ref{sec:flavorStructure}.

\subsection{Corrections to SM couplings}
In the SM the gauge couplings  $g$ and $g^\prime$, the electric charge $e$, the Fermi constant $G_F$, the Weinberg angle $\sin\theta_W$,
the gauge boson masses,  and the Higgs boson mass and $vev$ are not independent, but are connected by several relations, which hold at tree level, and are modified in a testable way by radiative corrections.
Including the higher dimensional operators of Eqs.\ \eqref{eq:Xphi2}, \eqref{eq:dipole}, \eqref{eq:vertex} and \eqref{eq:fourfermion}
modifies these relations, making it important to specify the input that is taken from the experiments.
Here we use as experimental input the values of the weak boson and Higgs boson masses $m_Z$, $m_W$, $m_H$ \cite{Olive:2016xmw},  $G_F$, the electric charge, and $\sin\theta_W$ as extracted from the lepton asymmetries at LEP \cite{Olive:2016xmw}. 
In the presence of new physics, these experimental values would include contributions from dimension-six operators, which we now describe.

The operators in Eq.\ \eqref{eq:Xphi2} affect the normalization of the gauge bosons, and their masses, inducing, in particular, mass and kinetic mixing between $W^3_\mu$ and $B_\mu$.
Here we follow Ref.\ \cite{Buchmuller:1985jz}, and diagonalize the gauge boson mass terms up to corrections of $\mathcal O(v^4/\Lambda^4)$, by 
rotating the $W_3$ and $B_\mu$ fields according to 
\begin{eqnarray}
W^3_\mu &=& \sw (1 + \alpha_{AA} ) A_\mu + \left( \cw (1 + \alpha_{ZZ}) - \sw \alpha_{AZ}  \right) Z_\mu ,\\
B_\mu   &=& \cw (1 + \alpha_{AA} ) A_\mu - \left( \sw (1 + \alpha_{ZZ}) + \cw \alpha_{AZ}  \right) Z_\mu,
\end{eqnarray}
where we have introduced
\begin{eqnarray}\label{rotation}
\alpha_{AA} &=& \swsq  \, C_{\varphi W} + \cwsq\, C_{\varphi B} - \sw \cw \, C_{\varphi W B} ,\\
\alpha_{ZZ} &=& \cwsq  \, C_{\varphi W} + \swsq\, C_{\varphi B} + \sw \cw \, C_{\varphi W B} ,\\
\alpha_{AZ} &=& 2 \sw \cw ( C_{\varphi B}- C_{\varphi W} ) +  \left( \cwsq - \swsq \right) \, C_{\varphi W B}. \label{rot3}
\end{eqnarray}
The subscript $0$ has been introduced to  stress that Eqs.\ \eqref{rotation}-\eqref{rot3} involve the Weinberg angle in the SM 
\begin{equation}
\cw = \cos \theta_W^0  =  \frac{g}{\sqrt{g^2 + g^{\prime\, 2}}}.
\end{equation}
In a similar way, we redefine $W^{1,2}_\mu$ to ensure the kinetic term of the $W$ boson is canonically normalized. 
The weak boson masses are then given by Ref.~\cite{Buchmuller:1985jz}
\begin{eqnarray}
m_W &=& m^0_{W} \left( 1 + C_{\varphi W} \right), \qquad  m_W^0 = \frac{g v}{2} ,\\
m_Z &=& m^0_{Z} \left( 1 + \alpha_{ZZ} \right),  \qquad  m_Z^0 = \frac{g v}{2 \cw}, 
\end{eqnarray}
for which we use the experimental values $m_W = 80.4$ GeV and $m_Z = 91.2$ GeV \cite{Olive:2016xmw}.
Notice that here we focused on the subset of the complete dimension-six Lagrangian presented in the previous section. The full expression of the $W$ and $Z$ boson masses in the SM-EFT are given in Ref.\ \cite{Buchmuller:1985jz,Alonso:2013hga}.

The Higgs $vev$ is defined in term of the observed Fermi constant. Up to corrections of $\mathcal O(v^4/\Lambda^4)$, the operators in Eqs.\ \eqref{eq:Xphi2}, \eqref{eq:dipole}, \eqref{eq:vertex} and \eqref{eq:fourfermion}
do not affect the relation between $v$ and $G_F$, and we still have
\begin{eqnarray}
v =  (\sqrt{2} G_F)^{-1/2}  = 246.2 \, \textrm{GeV}, \label{eq:vev}
\end{eqnarray}
where $G_F$ is the measured Fermi constant, extracted from muon decay, $G_F = 1.166 \cdot 10^{-5}$ GeV$^{-2}$.
Within the general SM-EFT the above relation receives corrections from additional operators  that we do not consider here \cite{Buchmuller:1985jz}. For example, at dimension-six 
Eq.\ \eqref{eq:vev} is altered by the operator $c^{(3)}_{L\varphi}$ in Eq.\ \eqref{eq:vertex}, which modifies the leptonic couplings of the $W$.
Using the physical masses and $G_F$, we define the effective couplings 
\begin{equation}
g_{\rm eff} = 2 \frac{m_W}{v} = 0.653, \qquad   \left(\frac{g}{c_w}\right)_{\rm eff} = 2 \frac{m_Z}{v} = 0.741,
\end{equation}
which are useful in expressing the  couplings of the quarks to the $W$ and $Z$ bosons.
The electric charge is given by Ref.~\cite{Olive:2016xmw}
\begin{equation}
e = g \sw \left(1 + \alpha_{AA}\right), \qquad \al(m_Z) = \frac{e(m_Z)^2}{4\pi}=1/127.95\,.
\end{equation}

Finally, we  use the lepton asymmetry parameter $A_e$, as measured at LEP \cite{Olive:2016xmw}, to determine the Weinberg angle. 
$A_e$ is related to the ratio of the electron vector and axial coupling, and, since we are not introducing new $\bar{e} e Z$ interactions, this ratio is only affected by the definition of the $Z$ field, 
\begin{equation}
\frac{g_V^e}{g_A^e} = 1 - 4 |Q_e| \swsq \, \left( 1 + \frac{\cw}{\sw} \alpha_{AZ}  \right).
\end{equation}
Using $g_V^e/g_A^e = 0.151 \pm 0.002$ \cite{Olive:2016xmw},  we define an effective $s_w$
\begin{equation}
(s^2_w)_{\rm eff} =  \swsq \, \left( 1 + \frac{\cw}{\sw} \alpha_{AZ}  \right) = 0.231.
\end{equation}

With the above definitions, the SM couplings of the $Z$ and $W$ boson to quarks can be written as
\begin{eqnarray}
\mathcal L & = & - \left(\frac{g}{c_w} \right)_{\rm eff}  \Bigg\{  \bar q_L \left(  T_3  - Q (s^2_w)_{\rm eff}   \right) \gamma^\mu q_L\, Z_\mu  - Q_u (s^2_w)_{\rm eff} \, \bar u_R \gamma^\mu  u_R\, Z_\mu \nonumber  \\
& & -  Q_d (s^2_w)_{\rm eff}\, \bar d_R \gamma^\mu d_R \, Z_\mu   \Bigg\} - \frac{(g)_{\rm eff}}{\sqrt{2}} \bar u_L \gamma^\mu d_L W^+_\mu, 
\end{eqnarray}
where $T_3$ is the third component of isospin, $Q = {\rm diag}(Q_u,Q_d)$ is the quark charge matrix.
The couplings of the $Z$ and $W$ boson to the Higgs boson are given by
\begin{eqnarray}
\mathcal L &=&  \frac{m^2_W}{v} h \left( W^{\mu}_1 W_{1\,\mu} + W^{\mu}_2 W_{2\,\mu}  \right)  + \frac{m^2_Z}{v} h Z^{\mu} Z_\mu.
\end{eqnarray}
We will compute the effects of dimension-six operators with respect to these definitions.

Lastly, the Yukawa operators in Eq.~\eqref{eq:yuk} affect the definition of the quark masses. We define the Yukawa Lagrangian in the presence of dimension-four and dimension-six operators as
\begin{equation}
\mathcal L = - v \bar u_L Y_u  u_R   \left(1 + \frac{h}{v}\right) - v \bar d_L Y_d  d_R   \left(1 + \frac{h}{v}\right) - \bar u_L Y_u^\prime u_R h,
  - \bar d_L Y_d^\prime d_R h + \textrm{h.c.}
\end{equation}
where 
\begin{equation}
Y_u  =  Y_u^{0}  +   \frac{1}{2} Y_u^\prime, \qquad  Y_d =  Y_d^{0}  +   \frac{1}{2} Y_d^\prime,
\end{equation}
and $Y^0_{u,d}$ denote the Yukawa couplings in the SM.
The matrices $Y_u$ and $Y_d$ can always be made diagonal and real by appropriate field redefinitions, and are determined by the observed quark masses, $M_{u,d} = v Y_{u,d}$.   
In the mass basis,  $Y^\prime_{u,d}$ are generic $3\times 3$ complex matrices.
In what follows we will neglect the contribution of $Y_{u,d}$, which,  for processes involving light quarks, is always negligible, and we will focus on $Y_{u,d}^\prime$, 
which, being unrelated to the light quark masses, need to be independently constrained.

\subsection{Flavor assumptions}\label{sec:flavorStructure}

In the code we try to keep the flavor structure of the couplings in Eqs.\ \eqref{eq:dipole}, \eqref{eq:vertex} and \eqref{eq:fourfermion} as generic as possible. 
In the lepton sector, we work under the assumption of lepton universality, which 
implies that the four-fermion and leptonic dipole operators in Eqs.\ \eqref{eq:fourfermion} and \eqref{eq:dipole} have a trivial leptonic structure,
with the same couplings to the electron, muon, and tau.  
This assumption can  easily be relaxed by setting the \texttt{vdecaymode} flag in the \texttt{powheg.input} file to separately only select electron, muon or tau final states. 
At the moment the code does not allow for lepton-flavor-violating couplings. In this paper we neglect neutrino masses and define the operators in the neutrino flavor basis.

To define the couplings to quarks, we first rotate to the quark mass basis, by performing the following transformation
\bea
u_{L,R}\rightarrow U^u_{L,R}u_{L,R},\qquad d_{L,R}\rightarrow U^d_{L,R}d_{L,R}, \label{basistransf}
\eea
where $U^{u,d}_{L,R}$ are unitary matrices that diagonalize the mass matrices 
\bea
(U_L^u)^\dagger M_u U_R^u ={\rm diag}(m_u,\,m_c,\,m_t),\qquad 
(U_L^d)^\dagger M_d U_R^d ={\rm diag}(m_d,\,m_s,\,m_b),
\eea
and the CKM matrix is $V_{\rm CKM }= (U^u_L)^\dagger U^d_L$.
For most operators, the effects of the transformation can be trivially absorbed by a redefinition of the couplings.
For example, for the right-handed $W$ current $\xi$, the net effect of switching to the mass basis is to replace the matrix $\xi$ with $\xi^\prime = (U^u_R)^\dagger\,  \xi \, U_R^d$.
Since $\xi$ is an arbitrary matrix, the change has no practical effect. $c_{U\vp}$, $c_{D\vp}$, $Y_{u,d}^\prime$, $C_{eu}$, $C_{ed}$, $C_{Lu}$, $C_{Ld}$ only affect neutral currents, and we simply define them to be  
$3\times3$ matrices in the mass basis.  Such a redefinition does not work for $C_{Qe}$, which induces new neutral-current couplings for both $u$-type and $d$-type quarks. The Lagrangian in the mass basis is 
\begin{equation}
\mathcal L = - \frac{4 G_F}{\sqrt{2}} \bar e_R \gamma^{\mu} e_R \, \left( \bar u_L (U^{u}_L)^{\dagger} C_{Qe} U_L^{u} \gamma_\mu  u_L + \bar d_L (U^{d}_L)^{\dagger} C_{Qe} U_L^{d}\gamma_\mu  d_L \right).
\end{equation}
In this case we \textit{define} the coupling for $d$-type quarks, $C^{\prime}_{Qe} \equiv (U^{d}_L)^{\dagger} C_{Qe} U_L^{d}$, resulting in the appearance of factors of the CKM matrix for $u$-type couplings.
Dropping the prime, we get
\begin{equation}
\mathcal L = - \frac{4 G_F}{\sqrt{2}} \bar e_R \gamma^{\mu} e_R \, \left( \bar u_L V_{\rm CKM} C_{Qe} V_{\rm CKM}^{\dagger} \gamma_\mu u_L + \bar d_L  C_{Qe} \gamma_\mu d_L \right).
\end{equation}
As a consequence, we label the elements of $C_{Qe}$ with $d$-type indices in Table \ref{TabA}.

The rotation \eqref{basistransf} also has nontrivial consequences for the operators $c_{Q\vp}^{(1)}$, $c_{Q\vp}^{(3)}$, 
$\Gamma_W^u$, $\Gamma_W^d$, $C^{(1)}_{LQ}$, $C^{(3)}_{LQ}$, $C^{}_{LedQ}$,  $C^{(1)}_{LeQu}$, $C^{(3)}_{LeQu}$, which affect both neutral and charged currents.
In these cases, we will absorb the rotation matrices that appear in neutral-current interactions in the definition of the coefficients of effective operators, 
resulting in the appearance of explicit factors of the CKM matrix   in charged-current interactions.

For example, for $c_{Q\vp}^{(1)}$ and $c_{Q\vp}^{(3)}$ we find, in the unitary gauge and the weak basis,
\begin{eqnarray}
\mathcal L &=& - \left(1 + \frac{h}{v}\right)^2  \Bigg\{  \frac{g}{2 c_w}  \bar q_L \gamma^\mu Z_{\mu} \left(    c_{Q\vp}^{(1)}  - c_{Q\vp}^{(3)}\tau_3  \right) q_L   - \frac{g}{\sqrt{2}}   \bar u_L  \gamma^\mu W_{\mu}^+  \,  c_{Q\vp}^{(3)}  d_L  \nn \\ & & 
- \frac{g}{\sqrt{2}}  \bar d_L  \gamma^\mu W_{\mu}^-  \,  c_{Q\vp}^{(3)}  u_L \Bigg\}. 
\end{eqnarray}
Transforming to the mass basis, the left-handed couplings of the $Z$ boson to $u$-type and $d$-type quarks are
\begin{eqnarray}
 c_{Q\vp, \, U}  & =&   (U_L^u)^{\dagger}\, \left(  c_{Q\vp}^{(1)} -   c_{Q\vp}^{(3)}   \right) \,  U_L^{u} \qquad
  c_{Q\vp, \, D}   =  (U_L^d)^{\dagger}\,  \left(  c_{Q\vp}^{(1)} +   c_{Q\vp}^{(3)}   \right) \, U_L^{d}. \nonumber
\end{eqnarray}
These $Z$ couplings in the mass basis, $c_{Q\vp, \, U}$ and $c_{Q\vp, \, D}$, are $3\times3$ hermitian matrices.
As a consequence, the $W$ couplings are given by
\begin{eqnarray}
\mathcal L && = 
\frac{g}{2 \sqrt{2}} \left(1 + \frac{h}{v}\right)^2  \bar u_L \gamma^\mu W_{\mu}^+ \left(  V_{\textrm{CKM}}\,   c^{}_{Q\vp,\, D} - c^{}_{Q\vp,\, U} \, V_{\textrm{CKM}}       \right) d_L + \textrm{h.c.},
\end{eqnarray}
and are in general non-diagonal, even if one assumes that there are no tree-level flavor-changing neutral currents.

In a similar way, we define the $\Gamma_{W,\gamma}^{u,d}$ matrices to be the couplings of the $Z$ and $\gamma$ dipoles in the mass basis,  which results in the following couplings for the $W$ dipoles 
\bea \vL &=& -\frac{g}{\sqrt{2} v} \bigg(1+\frac{h}{v}\bigg) \Bigg\{ \bar d_L V_{\rm CKM}^\dagger \Gamma_W^u \simu u_R W_{\mu\nu}^- + \bar u_L V_{\rm CKM}\Gamma_W^d \simu d_R W_{\mu\nu}^+ \Bigg\}+{\rm h.c.}. 
\eea
Finally, the charged-current interactions mediated by semileptonic four-fermion operators are 
\begin{eqnarray}
\mathcal L &=& - \frac{4 G_F}{\sqrt{2}} \Bigg\{  \bar \nu_L \gamma^\mu e_L  \bar d_L \gamma_\mu \left( C_{LQ, D} V^{\dagger}_{\rm CKM} - V^{\dagger}_{\rm CKM} C_{LQ, U}  \right) u_L  \nn \\
& & + \bar \nu_L e_R \, \left( \bar d_R C_{LedQ} V^{\dagger}_{\rm CKM} u_L + \bar d_L V^{\dagger}_{\rm CKM} C^{(1)}_{LeQu} u_R  \right) +  \bar \nu_L \sigma^{\mu \nu} e_R \,   \bar d_L V^{\dagger}_{\rm CKM} C^{(3)}_{LeQu} \sigma_{\mu \nu} u_R  \Bigg\} \nonumber \\ & & + \textrm{h.c.}, 
\end{eqnarray}
where
\begin{eqnarray}
C_{LQ,\, U} &=&C^{(1)}_{LQ} - C^{(3)}_{LQ}, \nn \\ 
C_{LQ,\, D} &=& C^{(1)}_{LQ} + C^{(3)}_{LQ} ,
\end{eqnarray}
are the neutral-current couplings of these operators in the mass basis. 
With these definitions of the purely left-handed semileptonic operators, additional factors of $V_{\rm CKM}$ appear in the $p p \rightarrow \nu \bar{\nu}$ channel.

Hermiticity of the Lagrangian in Eqs. \eqref{eq:vertex} and \eqref{eq:fourfermion} forces the couplings $c_{U\vp}$, $c_{D\vp}$, $c^{}_{Q\varphi,\, U}$, $c^{}_{Q\varphi,\, D}$, $C_{eu}$, $C_{ed}$, $C_{Lu}$, $C_{Ld}$ and $C_{Qe}$  to be  $3\times3$ hermitian matrices. This implies that  CP-violating phases can only appear in off-diagonal elements. 
Since we do not expect collider probes to be competitive with low-energy precision experiments, at present these couplings are set to be real. CP-violating phases can be straightforwardly incorporated, if necessary.
The dipole couplings $\Gamma^{u,d,e}_{W,\gamma}$, the right-handed charge-current $\xi$, the Yukawa couplings $Y^\prime_{u,d}$, and the scalar and tensor interactions $C^{}_{LedQ}$, $C^{(1)}_{LeQu}$, $C^{(3)}_{LeQu}$ 
are generic $3\times3$ matrices. 
For these couplings, both the real and imaginary parts can be turned on. The notation for the coefficients of the operators that can
be set in the input file is listed in Table \ref{TabA}.

As an alternative to considering generic flavor structures, the code also provides the option to assume the Minimal Flavor Violation (MFV) \cite{D'Ambrosio:2002ex} framework. In this scenario, the allowed forms of the couplings are constrained by  flavor symmetries, which significantly decreases the number of free parameters. The implications for the flavor structures of the couplings, as well as the implementation of this scenario in the code, are discussed in Appendix \ref{app:MFV}.

In addition to the couplings of dimension-six operators, the user can input the elements of the CKM matrix by setting  $\texttt{CKM\_V}ij$, with $i \in \{\texttt{u,c,t}\}$ and $j \in \{ \texttt{d,s,b}\}$,
in the input file. By default the CKM matrix is assumed to be real, and terms of $\mathcal O(\lambda^2)$, where $\lambda \sim |V_{us}|$, and higher are neglected. 


\begin{table}
\center
\begin{small}
\begin{tabular}{||c|cc||c|cc||}
\hline
								  &  \multicolumn{2}{|c||}{Operator} 	 					&
								  &  \multicolumn{2}{c||}{Operator} 												 \\
\hline
$C_{\varphi W,\, \varphi B, \, \varphi W B}$, 			  & \multicolumn{2}{c||}{\texttt{CC\_ww}, \texttt{CC\_bb}, \texttt{CC\_wb} }      &   
$C_{\varphi \tilde{W},\varphi \tilde{B},\varphi \tilde{W} B}$  	  & \multicolumn{2}{c||}{ \texttt{CC\_wwt}, \texttt{CC\_bbt}, \texttt{CC\_wbt}   }     \\ 
\hline
$\Gamma^u_W$ 		& \texttt{ReGUw}\_\textit{ik}  				&   		&
$\Gamma^d_W$ 		& \texttt{ReGDw}\_\textit{jl}  				&    		\\%
			& \texttt{ImGUw}\_\textit{ik}  				&   		&
			& \texttt{ImGDw}\_\textit{jl}  				&    		\\
 $\Gamma^u_\gamma$   	& \texttt{ReGUe}\_\textit{ik} 				&   		& 
 $\Gamma^d_\gamma$   	& \texttt{ReGDe}\_\textit{jl}				&  		\\
			& \texttt{ImGUe}\_\textit{ik}   			&   		& 		    
			& \texttt{ImGDe}\_\textit{jl} 				& 	 	\\
$\Gamma^e_W$ 		& \multicolumn{2}{c||}{\texttt{ReGEw}, \texttt{ImGEw}} 		         & 
$\Gamma^e_\gamma$ 	& \multicolumn{2}{c||}{\texttt{ReGEe}, \texttt{ImGEe}} 					\\
\hline
$c^{}_{Q\varphi,\, U}$ 	& \texttt{QphiU}\_\textit{ik}				&   $i \ge k$ 	& 
$c^{}_{Q\varphi,\, D}$ 	& \texttt{QphiD}\_\textit{jl} 				&   $j \ge l$  	\\
$c_{U\varphi}$ 		& \texttt{Uphi}\_\textit{ik}	 			&   $i \ge k$ 	& 
$c_{D\varphi}$ 		& \texttt{Dphi}\_\textit{jl}	    			&   $j \ge l$  	\\
$\xi$          		&  \texttt{ReXi}\_{\textit{ ij}}  			&   & & 	&\\ 
			&  \texttt{ImXi}\_\textit{ij}     			&   & & 	&  \\
\hline
$Y^\prime_u$ 		& \texttt{ReYu}\_\textit{ik}  				&    	&
$Y^\prime_d$ 		& \texttt{ReYd}\_\textit{jl}				&    	\\ 
			& \texttt{ImYu}\_\textit{ik}  				&    	& 
			&\texttt{ImYd}\_\textit{jl} 				&    	\\
\hline
$C_{LQ,U}$ 		&  \texttt{QLu}\_\textit{ik} 				&   $i\ge k$ 	&  
$C_{LQ,D}$ 		&  \texttt{QLd}\_\textit{jl}				&   $j\ge l$ 	\\
$C_{eu}$   		& \texttt{Ceu}\_\textit{ik}  				&   $i\ge k$  	&
$C_{ed}$   		& \texttt{Ced}\_\textit{jl}				&   $j\ge l$  	\\
$C_{Lu}$   		& \texttt{CLu}\_\textit{ik}  				&   $i\ge k$  	&  
$C_{Ld}$   		& \texttt{CLd}\_\textit{jl}				&   $j\ge l$    \\
$C^{(3)}_{LeQu}$   	&  \texttt{ReLeQu3}\_\textit{ik}  			&       	&  
$C_{Qe}$ 		& \texttt{Qe}\_\textit{jl} 				&   $j\ge l$ 	\\
			& \texttt{ImLeQu3}\_\textit{ik}		       		&     &  	                            & &    \\
$C^{(1)}_{LeQu}$ 	& \texttt{ReLeQu}\_\textit{ik} 				&     & 
$C_{LedQ}$  		& \texttt{ReLedQ}\_\textit{jl} 				&   	\\
			& \texttt{ImLeQu}\_\textit{ik}	 			&     & 		       & \texttt{ImLedQ}\_\textit{jl} &    \\
\hline
\end{tabular}
\end{small}
\caption{Notation for the coefficients of the dimension-six operators that can be set in \POWHEG{}.
The flavor structures reflect the assumptions discussed in Section \ref{sec:flavorStructure}.
The indices $i, k$ run on $u$-type quark flavors $\textit{i},\textit{k} \in \{\texttt{u,c,t}\}$, while $j, l$ on $d$-type quark flavors $\textit{j},\textit{l} \in \{\texttt{d,s,b}\}$.
With the assumptions of Section \ref{sec:flavorStructure}, $c^{}_{Q\varphi,\, U}$, $c^{}_{Q\varphi,\, D}$, $c^{}_{U\varphi}$, $c^{}_{D\varphi}$, 
$C_{LQ,U}$, $C_{LQ,D}$, $C_{eu}$, $C_{ed}$, $C_{Lu}$, $C_{Ld}$ and $C_{Qe}$ are real symmetric matrices. The notation $i \ge k$ and $j \ge l$ indicates that the elements $\{\texttt{uu,uc,ut,cc,ct,tt}\}$ 
and $\{\texttt{dd,ds,db,ss,sb,bb}\}$ can be set by the user, while the remaining elements are not independent. }\label{TabA}
\end{table}

\subsection{Renormalization}

The coefficients of most operators in Eq.\ \eqref{eq:basis1} do not run in QCD at one loop. 
The exceptions are the quark Yukawa couplings in Eq.\ \eqref{eq:yuk} and  the  scalar operators in Eq.\ \eqref{eq:fourfermion}, whose coefficients $Y^\prime_{u,d}$,  $C_{LeQu}^{(1)}$ and $C_{LedQ}$ obey the same renormalization group equation (RGE) as the quark masses,
the  dipole operators in Eq.\ \eqref{eq:dipole}, and the tensor operator in Eq.\ \eqref{eq:fourfermion}. 
The scalar and tensor operators satisfy
\begin{eqnarray}\label{rge1}
\frac{d}{d \log \mu}  C_S &=&  \frac{\alpha_s}{4\pi} \sum_n  \left( \frac{\alpha_s}{4\pi} \right)^n \gamma^{(n)}_S
C_S, \qquad  C_S \in \left\{Y_u^\prime, Y^{\prime}_d, C^{(1)}_{LeQu}, C^{}_{LedQ}\right\}  \nonumber \\
\frac{d}{d \log \mu}  C_T &=&  \frac{\alpha_s}{4\pi} \sum_n  \left( \frac{\alpha_s}{4\pi} \right)^n \gamma^{(n)}_T
C_T, \qquad  C_T \in \left\{\Gamma_W^u, \Gamma_W^d, \Gamma_\gamma^u, \Gamma_\gamma^d, C^{(3)}_{LeQu}\right\}, 
\end{eqnarray}
where the two loop anomalous dimensions are \cite{Misiak:1994zw,Vermaseren:1997fq,Degrassi:2005zd} 
\begin{eqnarray}\label{rge2}
\gamma_S^{(0)} &=& - 6 C_F, \qquad  \gamma_S^{(1)} = - \left( 3 C_F  + \frac{97}{3}  N_C - \frac{10}{3} n_f \right) C_F  ,\nonumber \\
\gamma_T^{(0)} &=&   2 C_F, \qquad  \gamma_T^{(1)} = \left(\frac{257}{9}  N_C -19 C_F -\frac{26}{9} n_f\right) C_F .
\end{eqnarray}
Here $C_F = 4/3$, $N_C = 3$ and $n_f = 5$ is the number of light flavors.
All the other coefficients do not run in QCD, and we neglect electroweak loops. 

By default, all dimension-six corrections are switched off in the \POWHEG{} input card.
To investigate the effect of one or more dimension-six operator, the user  needs to set the flag \texttt{dim6} to 1 and specify the values of the dimensionless coefficients defined in Eqs.\ \eqref{eq:Xphi2}, \eqref{eq:dipole}, \eqref{eq:vertex} and  \eqref{eq:fourfermion}
in the input file. The notation for the coefficients of the operators that can
be set in the input file is listed in Table \ref{TabA}, while Table \ref{TabB} summarizes the processes we considered, and which operator can be turned on and off in each process. 
For all processes except VBF, the user can set flavor off-diagonal coefficients, which induce tree-level processes that are absent in the SM, such as $d \bar s \rightarrow e^+ e^-$ 
or $c \bar u \rightarrow h e^+ e^-$. To generate these processes, the flag \texttt{fcnc} should be set to 1 in the \POWHEG{} input card.

\begin{table}
\center
\begin{tabular}{||c||c|  c|  c|  c| c||c||}
\hline
	        & {$p p \rightarrow \ell \nu$}  & {$p p \rightarrow \ell^+ \ell^-, \, \nu \bar\nu$} & {$WH$} & {$ZH$} & {VBF}     \\
	        \hline
$ C_{\varphi W}$,$C_{\varphi \tilde W}$      &   		--	  & -- 	   	&  $\green{\checkmark}$  & $\green{\checkmark}$	     & $\green{\checkmark}$      \\
$ C_{\varphi B}$, $ C_{\varphi \tilde B}$    &   		--	  & -- 	 	&  --  & $\green{\checkmark}$	     & $\green{\checkmark}$      \\
$ C_{\varphi WB}$, $ C_{\varphi \tilde W \tilde B}$   &   -- 	  & --         	&  --   & $\green{\checkmark}$	     & $\green{\checkmark}$ \\
\hline 
$ \Gamma_\gamma^{u,d} $ 	&		$\green{\checkmark}$	  & $\green{\checkmark}$ 			    & --  & $\green{\checkmark}$	     & $\green{\checkmark}$    \\
$ \Gamma_W^{u,d} $ 	&		$\green{\checkmark}$	  & $\green{\checkmark}$ 			    & $\green{\checkmark}$  & $\green{\checkmark}$	     & $\green{\checkmark}$    \\
$ \Gamma_{\gamma,W}^{e} $ 	&		$\green{\checkmark}$	  & $\green{\checkmark}$ 			    & $\red{\times}$  & $\red{\times}$	     & --    \\
\hline 
$ c^{}_{Q\varphi, U}, c^{}_{Q\varphi, D}$ &		$\green{\checkmark}$	& $\green{\checkmark}$	  				 & $\green{\checkmark}$ &  $\green{\checkmark}$	    & $\green{\checkmark}$     \\
$ c_{U \varphi}, c_{D \varphi}$ & --     & $\green{\checkmark}$                        	         & -- & $\green{\checkmark}$	     & $\green{\checkmark}$     \\
$\xi$			&		$\green{\checkmark}$	&	-- 		 & $\green{\checkmark}$ & --  & $\green{\checkmark}$  \\ 
$ c^{(1,3)}_{L\varphi},\, c_{e \varphi}$ &  $\red{\times}$	& $\red{\times}$	 & $\red{\times}$	& $\red{\times}$	& -- \\
\hline
$Y_u^\prime$, $Y^\prime_d$ &  -- & -- & $\green{\checkmark}$ & $\green{\checkmark}$ & -- \\
\hline 
$C_{LQ, U}, C_{LQ,D}$  &		$\green{\checkmark}$	 & $\green{\checkmark}$ 			   & --  & 	-- & --       \\
$C_{e u}, C_{ed}$      &		--	& $\green{\checkmark}$ 			   & --  & 	-- & --       \\
$C_{L u}, C_{L d, Qe}$ &		--	& $\green{\checkmark}$ 			   & --  & 	-- & --       \\
$C_{LedQ}, C^{(1,3)}_{LeQu} $            &$\green{\checkmark}$	  & $\green{\checkmark}$ 			   & --   & -- 			    & --    \\
\hline
\end{tabular}
\caption{Contributions of the dimension-six operators in Eqs.\ \eqref{eq:Xphi2}, \eqref{eq:dipole}, \eqref{eq:vertex}, \eqref{eq:yuk} and \eqref{eq:fourfermion} to 
NC and CC Drell-Yan, associated production of a Higgs and a weak boson, and
Higgs boson production via vector boson fusion.
For each process, a $\green{\checkmark}$ indicates that the contribution has been implemented in \texttt{POWHEG},
a $\red{\times}$  that the operator contributes at tree level, but has 
been neglected because of the reasons explained in the text,
while a $-$ indicates that the operator does not contribute at leading order.}\label{TabB}
\end{table}

The new physics scale $\Lambda$, at which the coefficients are defined, can be specified 
by setting the flag $\texttt{LambdaNP}$ to the desired value. 
Eqs.\ \eqref{rge1} and \eqref{rge2} are then used to run the coefficients from $\Lambda$ to $\mu_R$, the renormalization scale of the  process of interest. By default, $\texttt{LambdaNP} = 1$ TeV.
For the coefficients that do not have QCD evolution, the flag $\texttt{LambdaNP}$ is irrelevant.

\section{Neutral- and charged-current Drell-Yan production}\label{Wprod}

We start by analyzing the contributions of the dimension-six SM-EFT operators to the processes $p p \rightarrow \ell^+ \nu_\ell$,  $\ell^-  \bar{\nu}_\ell$,
$\ell^+  \ell^-$ and $\nu\bar\nu$. 
The SM background to these processes is known with very high accuracy, including fixed next-to-next-LO (NNLO) QCD corrections \cite{Anastasiou:2003yy,Melnikov:2006kv,Catani:2009sm}
and NLO electroweak (EW) corrections \cite{Dittmaier:2001ay,Baur:2001ze,Baur:2004ig,CarloniCalame:2006zq,Dittmaier:2009cr}.
More recently, the interface of the NNLO predictions with the parton shower has been presented in Refs.~\cite{Karlberg:2014qua,Alioli:2015toa,Hoeche:2014aia}, and a quantitative assessment of the size of different QCD and EW corrections
has been discussed in Ref.~\cite{Alioli:2016fum}. 
Contributions from SM-EFT operators, at LO in QCD, have  been considered, for example,  in Refs.\ \cite{Cirigliano:2012ab,Brivio:2017btx,Brivio:2017vri,Farina:2016rws} and rescaled by the SM NNLO $K$-factor in Ref.~\cite{Greljo:2017vvb,Alioli:2017nzr}. 
In this work, we have calculated the NLO QCD corrections to the partonic processes mediated by SM-EFT operators in Table \ref{TabB}
and interfaced with the parton shower according to the \POWHEG{} method, extending the original work of Ref.~\cite{Alioli:2008gx}.

If the invariant mass of the leptons is close to the weak boson masses, NC and CC Drell-Yan production is dominated by the production of a $W$ or $Z$ boson, 
which subsequently decays into leptons. In this region, these processes are in principle sensitive to modifications of the SM $W$ and $Z$ couplings to  quarks,
represented by the operators $c^{}_{Q\varphi, U}$, $c^{}_{Q\varphi, D}$,  $c_{U \varphi}$ and $c_{D \varphi}$,  or to new interactions of the $W$ and $Z$ boson: the $W$ coupling to right-handed quarks, mediated by $\xi$, and the chiral-symmetry-breaking dipole couplings $\Gamma^{u,d, e}_{W, \gamma}$. For $W$ production, the contributions of 
$c^{}_{Q\varphi, U}$ and $c^{}_{Q\varphi, D}$ amount to a shift in the values of the CKM mixing matrix, which, in the presence of these operators, is no longer unitary. Instead, 
$\xi$ induces couplings to quarks with opposite chirality to the SM. 
If one considers observables that are symmetric under the exchange of the charged lepton and the neutrino momenta, the contributions of these operators are identical to those of the SM, making it difficult to obtain strong constraints \cite{Alioli:2017ces}. The right-handed nature of $\xi$ would manifest, for example, in a larger fraction of right-handed polarized $W$ bosons. Appreciable deviations, however, require large couplings that are already
ruled out by other collider and low-energy observables, such as associated production of a $W$ and a Higgs boson \cite{Alioli:2017ces}.
Similarly, in $Z$ production, the cross section induced by the operators $c^{}_{Q\varphi, U}$, $c^{}_{Q\varphi, D}$, $c_{U \varphi}$ and $c_{D \varphi}$
is qualitatively very similar to the SM, and thus other processes,  such as $e^+ e^- \rightarrow q \bar q$ at the $Z$ pole, or $p p \rightarrow H Z$, provide stronger bounds.
In the case of the dipole couplings $\Gamma_{W, \gamma}^{u,d,e}$, the cross section has a different shape, being enhanced for large values of $W$ boson transverse mass or the dilepton invariant mass. 
Furthermore, the different chiral structure of the vertex could be identified by looking at the $W$ and $Z$ boson polarization. 

When the dilepton invariant mass  is much larger than the $W$ or $Z$ boson mass, NC and CC Drell-Yan production receive important contributions from the four-fermion operators in Eq.\ \eqref{eq:fourfermion}.
$C_{LQ, U}$, $C_{LQ, D}$, $C_{LedQ}$ and $C_{LeQu}^{(1,3)}$ contribute to the neutral and charged-current processes, while the remaining operators in Eq.\ \eqref{eq:fourfermion} only contribute to neutral currents.
Most of the semi-leptonic interactions, namely $C_{LQ, U}$, $C_{LQ, D}$, $C_{eu}$, $C_{ed}$, $C_{L u}$, $C_{L d}$ and $C_{Qe}$, couple left- and right-handed quarks to left- or right-handed leptons, and therefore  modify the helicity structures which already exist in the SM.
On the other hand, $C_{LedQ}$ and $C^{(1)}_{LeQu}$ are scalar couplings, while  $C^{(3)}_{LeQu}$ has the form of a tensor coupling, all of which give rise to new helicity structures. 

We computed the $W$ and $Z$ production cross sections in the presence of the
dimension-six operators in Table \ref{TabB}, including NLO QCD corrections,
and implemented them in the \POWHEGBOXVTWO{}.
The cross section has the schematic structure
\begin{eqnarray}\label{schematic}
\sigma_{} &=& \sigma_{\textrm{SM}} +  \sum_i \, \sigma_{i} C_i +    \sum_{i j}  \sigma_{\, ij} C_{i} C_j,
\end{eqnarray}
where $\sigma_{}$ indicates a generic (differential) cross section in the $\ell \nu$,  $\ell^+ \ell^-$ or $\nu\bar\nu$ channels. 
If one neglects light-quark mass effects, only the semileptonic four-fermion operator  $C_{LQ, U}$ and $C_{LQ, D}$
and the vertex corrections $c^{}_{Q\varphi, U}$ and $c^{}_{Q\varphi, D}$ interfere with the SM amplitude  for $p p \rightarrow \ell \nu$, due to helicity considerations.  In the case of $Z$ production,
the index $i$ in the interference term in Eq.\ \eqref{schematic} runs over the operators 
\begin{equation}
\{c^{}_{Q\varphi, U}, c^{}_{Q\varphi, D}, c_{U\varphi}, c_{D\varphi}, C_{LQ,U}, C_{LQ,D}, C_{eu}, C_{ed}, C_{Lu}, C_{Ld}, C_{Qe}\} .\nn
\end{equation}
We included dimension-eight effects to guarantee the positivity of the cross section even for arbitrary large values of the couplings. For completeness, we also included interferences between different dimension-six operators.   The chiral structures of the effective operators strongly limit 
the number of such interference terms and in practice one needs only to consider the interference of the photon and $Z$ dipoles in $pp \rightarrow \ell^+ \ell^-$, and of the $u$-type scalar and tensor operators, $C^{(1,3)}_{LeQu}$, in $pp \rightarrow \ell^+ \ell^-$ and $pp \rightarrow \ell \nu_\ell$.

When extracting bounds on the coefficients of effective operators, one should make sure to be working in the regime of validity of the EFT. For the operators that interfere with the SM one can check that the bounds are not dominated by terms quadratic in the new physics couplings. 
For a clear discussion of this point, we refer to Refs.\ \cite{Contino:2016jqw,Brivio:2017vri}. 
For operators with different chiral structures than the  SM, the leading contribution to the $W$ and $Z$ production cross sections is quadratic in the BSM coupling, and goes as $\Lambda^{-4}$.
One might then worry that the constraints on these operators are not reliable because of missing $\Lambda^{-4}$ contributions from the interference of genuine dimension-eight operators with the SM. This concern is justified for bounds obtained from the total cross section, or from differential distributions that are not sensitive to the quark/lepton chiralities. However, dimension-six operators with different chiral structures than the SM leave clear signatures in observables such as the $W$ and $Z$ polarization fractions, that cannot be mimicked by dimension-eight operators. 
Thus, in the presence of deviations from the SM, it will be possible to disentangle   dimension-six operators with different chiral structure than the SM from genuine dimension-eight effects by studying more differential observables.

\begin{figure}
\includegraphics[width=0.96\textwidth]{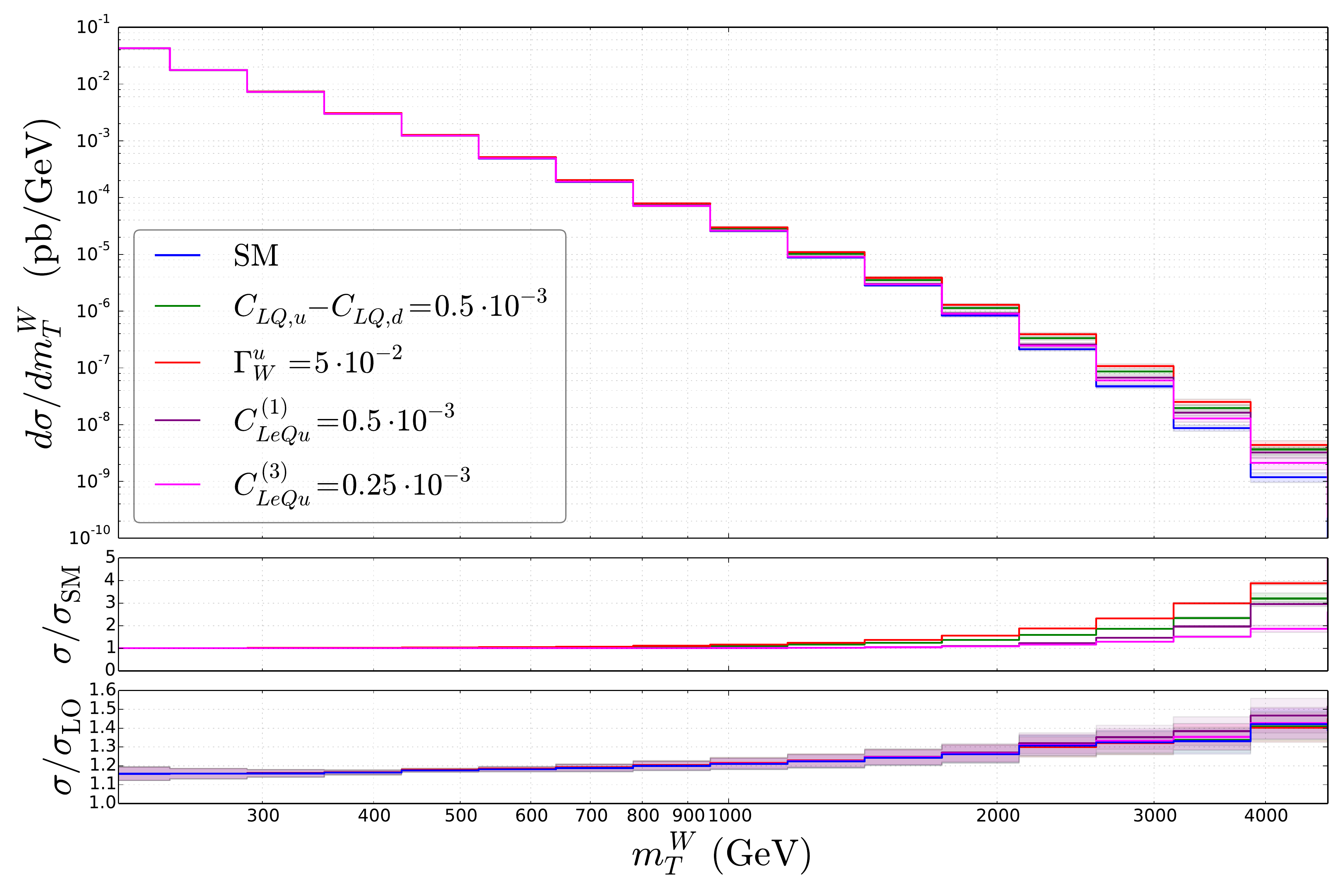}
\caption{Differential $p p \rightarrow e^+ \nu_e + e^- \bar\nu_e$ cross section as a function of $m^W_T$, at $\sqrt{S} = 13$ TeV. 
The middle panel shows the ratio of the differential cross sections in the presence of dimension-six operators and in the SM,
while the bottom panel the ratio of the NLO and LO cross sections.  The shaded regions indicate the theoretical uncertainties from PDF and scale variations. 
}\label{Fig:mtW}
\end{figure}

\begin{figure}
\includegraphics[width=0.96\textwidth]{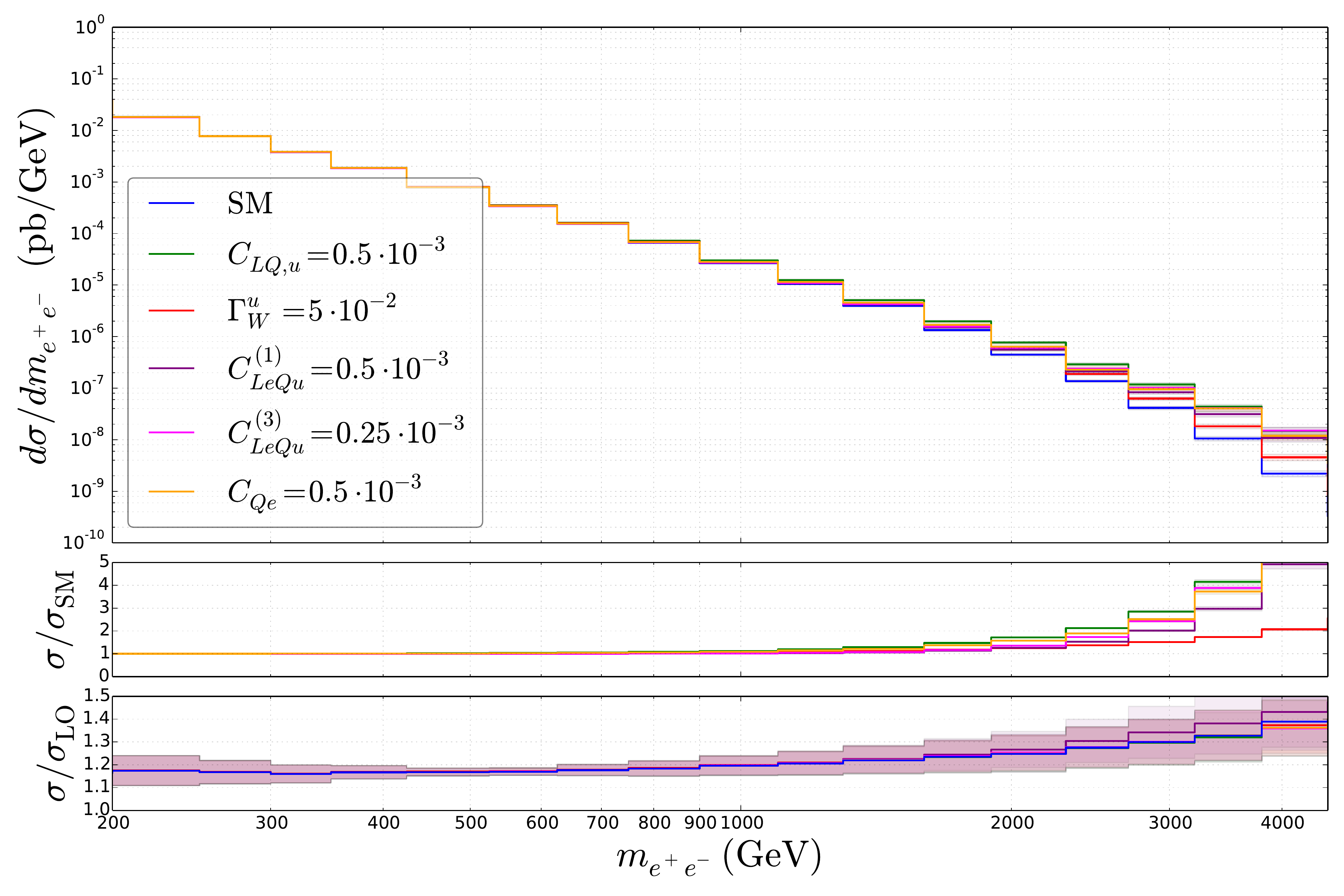}
\caption{Differential $p p \rightarrow e^+  e^-$ cross section as a function of the dilepton invariant mass, at $\sqrt{S} = 13$ TeV. The shaded regions indicate the theoretical uncertainties from PDF and scale variations.}\label{Fig:mZ}
\end{figure}


We show the resulting  $p p \rightarrow e^+ \nu_e + e^- \bar\nu_e$ cross section at $\sqrt{S} =13$ TeV
as a function of the $W$ transverse mass, $m^W_{T}$,  in  Fig.\ \ref{Fig:mtW}, where $m^W_{T} $ is defined as
\begin{equation}
m^W_{T} =  \sqrt{2 |p_{T}^\ell| |p_{T}^\nu| (1 - \cos \Delta \phi_{\ell\nu})}\, .
\end{equation}
Here $p_{T}^\ell$ and $p_{T}^\nu$ are the charged-lepton and the
neutrino transverse momenta, respectively,  and $\Delta \phi_{\ell\nu}$ is their
azimuthal separation.  
In blue we depict the SM cross section, while the remaining lines depict the contributions of the four-quark operators $C_{LQ, u} - C_{LQ, d}$ (green),  $C^{(1)}_{LeQu}$ (purple)  and $C^{(3)}_{LeQu}$ (magenta), and of the dipole operator $\Gamma_W^{u}$ (red). The values of the coefficients have been chosen to be close to the bounds discussed in Section \ref{bounds},
and only couplings to the first generation have been turned on.
The down-type and electron-type dipoles $\Gamma_W^{d}$, $\Gamma_W^{e}$, and $d$-type scalar operator $C_{LedQ}$
induce corrections with the same $m^W_T$ dependence as $\Gamma_W^u$ and $C^{(1)}_{LeQu}$, respectively.

In Fig \ref{Fig:mZ} we show $p p \rightarrow e^+ e^-$, as a function of the dilepton invariant mass. In addition to the operators shown in Fig.\ \ref{Fig:mtW}, we also included the coupling  of left-handed quarks to right-handed leptons, $C_{Q e}$, which does not contribute to $p p \rightarrow \ell \nu$. The photon dipole $\Gamma^{u}_\gamma$ gives a correction to the  cross section, which is roughly $70\%$ of the weak dipole at large $m_{e^+ e^-}$. The remaining $u$-type couplings, $C_{eu}$ and $C_{Lu}$, give rise to corrections to the cross sections that have similar shape as $C_{LQ, u}$ and $C_{Qe}$. Not shown in the figure are $d$-type couplings,
$\Gamma^d_W$, $\Gamma^d_\gamma$, $C_{LQ, d}$, $C_{ed}$, $C_{Ld}$, $C_{LedQ}$, which are qualitatively very similar to the corresponding $u$-type operators, with some suppression from the $d$-quark PDF.

The cross sections were evaluated using the \texttt{PDF4LHC15$\_$nlo$\_$30} PDF set \cite{Butterworth:2015oua}, with the factorization and renormalization scales set to $\mu_F = \mu_R = m_{\ell \ell^\prime}$,
where $m_{\ell \ell^\prime}$ is the invariant mass of the charged lepton and the neutrino, or of the two charged leptons. The running of the coefficients from the initial scale $\mu_0 = 1$ TeV to $\mu_R$
is taken into account by solving Eq.\ \eqref{rge1}.
The error bands in Figs.\ \ref{Fig:mtW} and \ref{Fig:mZ} include the $7-$point scale variations,
by independently varying $\mu_F$ and $\mu_R$ between $m_{\ell \ell^\prime}/2$ and $2
m_{\ell \ell^\prime}$ excluding the extremes, and PDF variations, computed with the 30 members of the \texttt{PDF4LHC15$\_$nlo$\_$30} PDF set.

For both $W$ and $Z$ production, the uncertainties of the NLO SM cross section are about 2-3\% at low $m^W_T$ or $m_{e^+ e^-}$, and increase to about 10\% at $m^W_T, m_{e^+ e^-} \sim 1-2$ TeV, where they are dominated by PDF uncertainties.
We find that the cross sections induced by the dimension-six operators that couple to the light quarks are affected by similar errors. 
In particular, the PDF uncertainties for both the SM and the dimension-six cross sections dominate at large $m^W_T$ or $m_{e^+ e^-}$, where they are about 10-15\%.  
The scale variations  for operators with a similar chiral structure as the SM,  such as $C_{L, Qu}$ or $C_{Qe}$,  as well as the dipole operators and the semileptonic tensor operators are all very similar, being at most around 5\%. The scalar operators $C_{LedQ}$ and $C^{(1)}_{LeQu}$, on the other hand, have larger scale uncertainties, close to $10\%$ at high invariant mass.


The cross section induced by the four-fermion  and  dipole operators,  as a function of $m^W_T$ or $m_{l^+ l^-}$, falls more slowly than in the SM, and thus the effects are more visible for large invariant mass. 
This is evident from the middle panels of Figs.\ \ref{Fig:mtW} and \ref{Fig:mZ}, which show the ratio of the differential cross sections in the presence of dimension-six operators and in the SM.
For the four-fermion operators that interfere with the SM, namely $C_{LQ, u}$, $C_{LQ, d}$, $C_{eu}$, $C_{ed}$, $C_{Lu}$, $C_{Ld}$ and $C_{Qe}$,  
the ratio scales as $(m^W_T/v)^2$ or $(m_{l^+ l^-}/v)^2$, whenever the interference term dominates.
The dipole operators  $\Gamma_{W,\gamma}^{u,d,e}$ do not interfere with the SM, and contribute at dimension-eight. In this case, the amplitude contains 
a $W$/$Z$ propagator, but it has an additional factor of momentum with respect to the SM, so that the squared amplitude is also enhanced by a factor of $(m_T^W/v)^2$.
Finally,  the scalar and tensor operators,  $C^{(1,3)}_{LeQu}$ and $C^{}_{LedQ}$, do not interfere with the SM and lack the $W$/$Z$ boson propagator, causing the ratio in the middle panel of Figs.\ \ref{Fig:mtW}
and \ref{Fig:mZ} to scale as 
$(m_T^W/v)^4$ and $(m_{\ell^+ \ell^-}/v)^4$ at high vector boson transverse or invariant mass.  

The bottom panels of Figs.\ \ref{Fig:mtW} and \ref{Fig:mZ} show the ratio of the NLO and LO Drell-Yan cross sections. 
We see that for both the SM and the dimension-six operators the importance of the NLO corrections increases at high $m_T^W$ or $m_{\ell^+ \ell^-}$, being around 30\%-40\% in the highest bins. 
Most dimension-six operators exhibit a behavior similar to the SM, with the scalar semileptonic operators $C_{LedQ}$ and $C^{(1)}_{LeQu}$ receiving the largest NLO corrections.
Since the contribution of dimension-six operators is particularly relevant at high invariant mass, the bottom panel of Fig.\  \ref{Fig:mtW} and \ref{Fig:mZ}
highlights the importance of including NLO QCD corrections\footnote{In the same high invariant mass region, higher-order EW corrections also become sizable and need to be taken into account for accurate predictions. }.

\subsection{Bounds on effective operators}\label{bounds}

We can take advantage of the enhancement at high $m_T^W$ and $m_{\ell^+ \ell^-}$ by interpreting the ATLAS and CMS searches for new high-mass phenomena in the dilepton  
or lepton and missing energy  final states \cite{Chatrchyan:2013lga,Aad:2014cka,Aad:2014wca,ATLAS:2014wra,Khachatryan:2014fba,Khachatryan:2016jww,Khachatryan:2016zqb,Aaboud:2017efa,Aaboud:2017buh} as bounds on the coefficients of the effective operators.   As an example of the kind of constraints one can obtain, we consider the analyses of Refs.\ \cite{Aaboud:2017buh} and \cite{Aaboud:2017efa}, which used 
data collected by the ATLAS collaboration during the LHC Run II at 13 TeV, with integrated luminosity of 36.1 fb$^{-1}$.  We calculated the $p p \rightarrow e^+ \nu_e$, $p p \rightarrow e^- \bar{\nu}_e$
and $p p \rightarrow e^+ e^-$ differential cross sections, interfacing to \PythiaEight\texttt{.219} \cite{Sjostrand:2007gs,Sjostrand:2014zea}
parton shower and hadronization, using the \POWHEG{} method. 
We used the same binning as the ATLAS collaboration, which   
considers 7 bins in $m_T^W$, between 130 GeV and 7 TeV  \cite{Aaboud:2017buh}, and 10 bins in $m_{e^+ e^-}$, from $m_Z$ up to 6 TeV \cite{Aaboud:2017efa}.
We applied the same cuts as the experimental collaboration on the leptons and missing energy, but did not perform a detailed detector simulation to determine acceptance and efficiency. 
In addition, we did not simulate additional SM backgrounds, such as $t \bar t$ and diboson production, but used the expected background events listed in Refs.\   
\cite{Aaboud:2017buh} and \cite{Aaboud:2017efa}.
For these reasons our bounds have to be interpreted as a crude estimate of the reach of the LHC.


\begin{table}
\center
\begin{tabular}{|c|cc||cc||cc|}
\hline
& \multicolumn{2}{c||}{LO $\times 10^{-4}$}  	 & \multicolumn{2}{c||}{NLO $\times 10^{-4}$}  &  \multicolumn{2}{|c|}{$\Lambda$ (TeV)} \\
\hline
$\left|\Gamma^u_W\right|$            	&	$< 480 $	   	& $< 710$	&  	$< 460$   	&  $< 640$ 	&    	1.1    	& 0.9	\\   
$\left|\Gamma^u_\gamma\right|$       	& 	$< 1200$ 		& $<1900$  	&       $<1200$   	&  $<1900$ 	& 	0.7    	& 0.6	\\   
$\left|\Gamma^d_W\right|$            	& 	$<510$ 			& $<730$	&   	$<470$     	&  $<650$ 	& 	1.1     & 1.0	\\
$\left|\Gamma^d_\gamma\right|$       	& 	$<1900 $		& $<2700$	&   	$<1800$ 	&  $<2600$ 	&  	0.6     & 0.5	\\
$\left|C_{LedQ}\right|$              	& 	$<6.9 $  		&  $<12$	& 	$< 5.7$  	&  $<10$ 	& 	10\,    &  7.7	\\
$|C^{(1)}_{LeQu}|$       	     	& 	$<4.6 $ 		&  $<10$	& 	$<3.9$    	&  $<8$ 	& 	12\,  	& 8.7	\\
$|C^{(3)}_{LeQu}|$       	     	& 	$<1.8 $  		&  $<4.4$	& 	$<1.6$  	&  $<4$   	&	19\,    &  12\,	\\
  $C_{LQ,u}$   			     	& 	$[-7.4,1.4] $ 		&  $[-16,3.8]$	&   	$[-6.9,1.2]$   	&  $[-14,3.0]$ 	&   	9.3   	& 6.6	 \\
  $C_{LQ,d}$	 		     	&  	$[-7.1,\,11] $		&  $[-14,\,20]$ &  	$[-6.9,9.5]$   	&  $[-13,\,17]$ &   	8.0 	&	6.0\\
 $C_{e,u}$    				&  	$[-6.9,3.4]$		&  $[-16,8.7]$	&   	$[-6.4,3.1]$    &  $[-14,7.7]$ 	&  	9.7 	& 6.6	\\
 $C_{e,d}$    				&  	$[-8.9,\,10] $		&  $[-17,\,20]$	 &  	$[-8.1,9.1]$   	&  $[-15,17]$ 	&  	8.1 	&	5.9\\
 $C_{L,u}$	 			&	$[-6.0,4.5]$		&  $[-14,\,11]$	 &   	$[-5.5,4.0]$   	&  $[-12,10]$ 	& 	10\,	& 7.1	\\
 $C_{L,d}$	 			&  	$[-9.0,9.9]$		&  $[-18,\,19]$	&   	$[-8.4,8.5]$   	&  $[-15,16]$ 	&   	7.8 	&	6.1\\
 $C_{Q,e}$	 			&  	$[-5.0,4.0]$		&  $[-11,\,10]$	&   	$[-4.8,3.4]$   	&  $[-9.6,8.6]$ &   	11\,	& 8.2	\\
   \hline
\end{tabular}
\caption{90\% CL bounds on the coefficients of SM-EFT operators that contribute to CC and NC Drell-Yan production, and the corresponding estimates of the scale $\Lambda$, assuming $C_i \equiv v^2/\Lambda^2$.
The bounds on the first and second columns use, respectively, SM-EFT cross sections at LO and NLO in QCD. 
$\Lambda$ is extracted from the NLO bounds, and, for asymmetric limits,  the weaker limit is used.  
The bounds are obtained by turning on all operators at the scale $\mu_0 = 1$ TeV, but with the assumptions that only the couplings to the $u$ and $d$ quarks are non-zero.  
In each column, the first bound uses all the bins in Refs.\ \cite{Aaboud:2017buh} and \cite{Aaboud:2017efa}, corresponding to a maximum $m_T^W$ and $m_{e^+ e^-}$ of 
$(m_T^W)_{\rm max} = [3,7]$ TeV and $(m_{e^+ e^-})_{\rm max} = [3,6]$ TeV, while 
the second bound excludes the last bin, corresponding to  $(m_T^W)_{\rm max} = [2,3]$ TeV and $(m_{e^+ e^-})_{\rm max} = [1.8,3]$ TeV.
}\label{WZbounds}
\end{table}

The resulting 90\% confidence limits are shown in Table \ref{WZbounds}. We turn on all operators at the new physics scale $\mu_0 =$ 1 TeV, assuming that only the couplings to the first generation 
of quarks are non-zero,
and use the electron channels  of Refs.\ \cite{Aaboud:2017buh} and \cite{Aaboud:2017efa}, 
thus constraining  the electron component of four-fermion operators. 
The muon component can be analyzed in a similar fashion. The bounds in the first column of Table \ref{WZbounds} are extracted using a LO calculation of the correction to the cross section from dimension-six operators, while those on the second column include NLO QCD corrections.
We show the limits in two cases. The first bound in each column  includes all the bins of Refs.\ \cite{Aaboud:2017buh} and \cite{Aaboud:2017efa}. In this case, the limits are dominated by the last bin
in $m^W_T$, $m^W_T \in [3,7]$ TeV, and $m_{e^+ e^-}$, $m_{e^+ e^-}  \in [3,6]$ TeV. The second bound  excludes the highest bin, probing 
$m_T^W$ and $m_{e^+ e^-}$ up to 3 TeV.

From Table \ref{WZbounds}, one can see that the processes considered in this section are not very sensitive to the dipole couplings $\Gamma^{u,d}_{W,\gamma}$. 
These coefficients are constrained at the 10\% level. If we translate these bounds to an estimate of the new physics scale $\Lambda$, using $\Gamma^{u,d}_{W,\gamma} \equiv v^2/\Lambda^2$, we
see that the effective scale would be $\lesssim 1$ TeV. This is smaller than the scales directly accessed by the experiment, implying the EFT framework might not be applicable in this region and the resulting limits should be interpreted with some caution.
On the other hand, the semileptonic four-fermion operators are strongly constrained, in several cases at better than the permil level, corresponding to  $\Lambda \sim 10$ TeV.
For scalar and tensor operators, the strong bounds suggest that the last two bins are in the regime of validity of the EFT, $\Lambda \gg m^W_T$, $m_{e^+ e^-}$.
For the remaining operators we checked that for the values of the couplings in Table \ref{WZbounds} the interference term, linear in $v^2/\Lambda^2$, is larger than the quadratic piece.

Comparing the first and second column of Table \ref{WZbounds}  we see that including NLO QCD corrections impacts the coefficients of dimension-six operators at the 10\% - 20\% level. 
As expected from Figs.\ \ref{Fig:mtW} and \ref{Fig:mZ}, the effect is larger for the analysis that uses all bins of Refs.\ \cite{Aaboud:2017buh} and \cite{Aaboud:2017efa}, which is dominated by the bin of highest invariant mass, and the scalar operators $C_{LedQ}$
and $C_{LeQu}^{(1)}$ are the most affected by NLO corrections.

It is interesting to compare the constraints of Table \ref{WZbounds} with
complementary constraints that can be extracted from low-energy observables.
First, we point out that the imaginary parts of several operators can be stringently constrained by electric-dipole-moment (EDM) measurements.
For example, the imaginary parts of $C^{(1)}_{LeQu,LedQ}$ and $C^{(3)}_{LeQu}$ are probed by measurements of $T$-violation in paramagnetic systems and in diamagnetic atoms, respectively. Below the QCD scale, the scalar and tensor couplings induce spin-independent and spin-dependent electron-nucleon interactions, often referred to as $\tilde C_S$ and $\tilde C_T$, respectively \cite{Fleig:2018bsf,Jung:2013hka,Engel:2013lsa}. The scalar coupling is most sensitively probed in measurements of $T$-violation in ThO, while the EDM of $^{199}$Hg is sensitive to $\tilde C_T$. The current experimental constraints \cite{Baron:2013eja,Griffith:2009zz} roughly imply ${\rm Im}\,C^{(1)}_{LeQu,\,LedQ}\lesssim 10^{-9}$ and ${\rm Im}\,C^{(3)}_{LeQu}\lesssim 10^{-10}$.
Similarly, the imaginary parts of the dipole couplings  also induce EDMs at low energies. The resulting  limits are especially stringent for the couplings to the photon, $\Gamma_{\gamma}^{u,d}$, which  directly contribute to the neutron EDM \cite{Bhattacharya:2015esa,Bhattacharya:2015wna}. Instead, the $\Gamma_{W}^{u,d}$ couplings induce the neutron EDM at the one-loop level \cite{Alonso:2013hga}. The current experimental constraints \cite{Baker:2006ts,Baron:2013eja,Afach:2015sja} give Im  $\Gamma_{W}^{u,d}\lesssim 10^{-6}$ and Im $\Gamma_{\gamma}^{u,d}\lesssim 10^{-9}$. 

The low-energy constraints that can be set on the real parts of the couplings are weaker, and comparable to the direct limits in Table \ref{WZbounds}.
Focusing on the four-fermion operators $C^{(1,3)}_{LeQu}$ and $C^{}_{LedQ}$, we can map them into the  scalar, pseudoscalar, and tensor couplings introduced in Refs.\ \cite{Cirigliano:2012ab,Cirigliano:2013xha}, 
$\epsilon_S^* =  C^{(1)}_{LeQu} + C_{LedQ}$, $\epsilon_P^* =  C^{(1)}_{LeQu} - C_{LedQ}$ and $\epsilon_T^* = C^{(3)}_{LeQu}$. 
The pseudoscalar coupling $\epsilon_P$ is strongly bound by the leptonic decay of the pion, and, in particular, by the ratio $\Gamma(\pi \rightarrow e \nu)/\Gamma(\pi \rightarrow \mu \nu)$,
which is suppressed by $m^2_e/m^2_\mu$ in the SM. $\epsilon_S$ is constrained by superallowed $\beta$ decays \cite{Hardy:2014qxa}, while the tensor coupling $\epsilon_T$ affects radiative pion decays,
such as $\pi \rightarrow e \nu \gamma$ and the neutron decay parameters \cite{Cirigliano:2009wk,Cirigliano:2012ab,Cirigliano:2013xha}. The analysis of Refs.\ \cite{Cirigliano:2009wk,Cirigliano:2012ab,Cirigliano:2013xha,Gonzalez-Alonso:2018omy} shows that the bounds on the
pseudoscalar coupling  are at the level $|\epsilon_P| \lesssim 5 \cdot 10^{-4}$, while for the scalar and tensor $|\epsilon_{T,S}| \lesssim  10^{-3}$, thus making collider observables very competitive with low-energy probes for the real parts of these couplings.

\begin{figure}
\center
\includegraphics[width=0.495\textwidth]{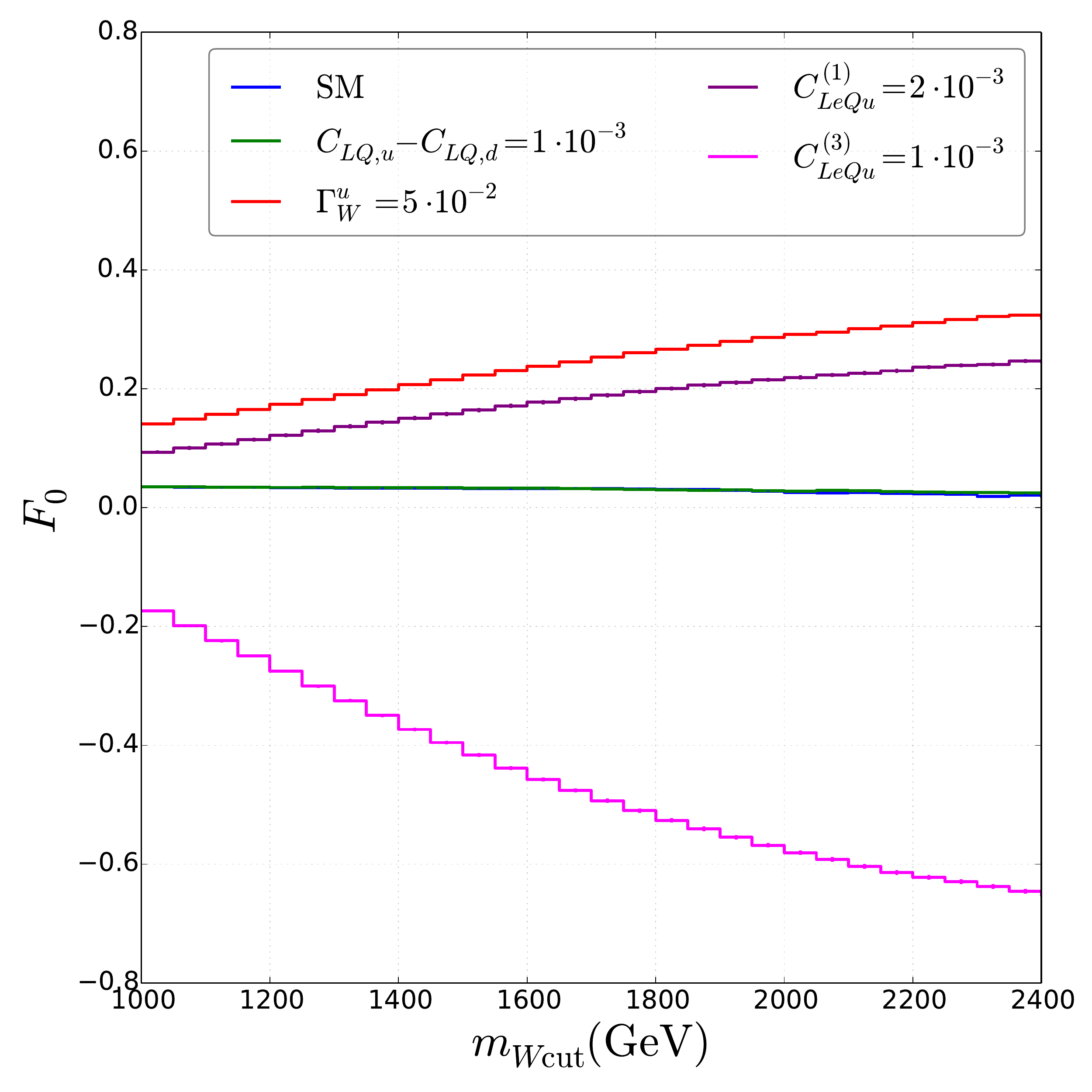}
\includegraphics[width=0.495\textwidth]{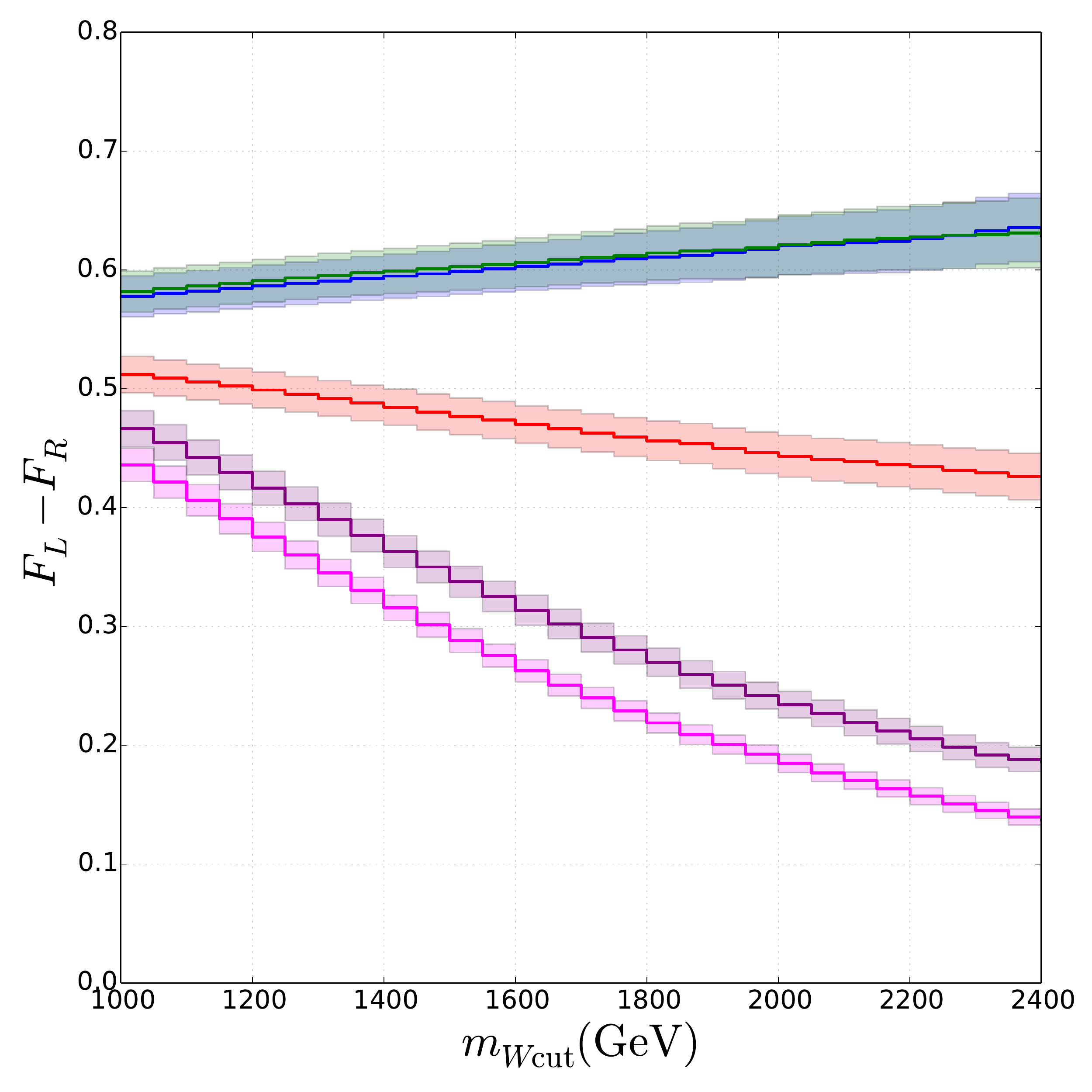}
\caption{Polarization fractions $F_0$ and $F_L - F_R$ in $pp \rightarrow e^+ \nu_e$, for $m_W > m_{W\, \rm cut}$, where $m_W$ denotes the invariant mass of the neutrino and charged lepton, at $\sqrt{S} = 13$ TeV.
The shaded regions indicate the uncertainties from PDF and scale variations.
}\label{Fig:W2}
\end{figure}


\subsection{Disentangling dimension-six operators}
In light of the comparable sensitivities of low- and high-energy observables, it is very important to construct collider observables that, in the presence of deviations from the SM, are able to differentiate between the various  dimension-six operators.
Examples are angular distributions. For $p p \rightarrow \ell \nu_\ell$ we work in the $W$ helicity frame, that is the frame 
in which the 
$\ell \nu_\ell$ system is at rest, with the $z$-axis chosen to be in the direction of the sum of the momenta of the charged lepton and neutrino in the laboratory frame.
We consider the differential distribution with respect to $\cos \theta^*$, where $\theta^*$ is the polar angle of the charged lepton in this  frame. 
When the charged lepton and the neutrino emerge from the decay of a $W$  boson, 
the differential  $\cos \theta^*$ distribution is related to the $W$ boson polarization fractions, $F_{0}$, $F_L$, and $F_R$ \cite{Bern:2011ie,Stirling:2012zt}.
We can write the differential distribution as 
\begin{eqnarray}\label{eq:angles}
\frac{1}{\sigma} \frac{d\sigma}{d\cos\theta^*} &=& \frac{3}{8} \left[  1 + \cos^2\theta^*  + \frac{A_0}{2} (1- 3 \cos^2 \theta^*)  + A_4 \cos\theta^* + \ldots \right],
\end{eqnarray}
where the dots denote higher powers of $\cos\theta^*$, which can arise if the 
$\ell \nu_\ell$ system has total angular momentum $J > 1$. $\sigma$ denotes any differential cross section that does not depend on the kinematics of individual leptons.
If the process is mediated by a $W$ boson, the $W$ polarization fractions are given by
\begin{equation}\label{F0LR}
F_0 = \frac{A_0}{2}\, , \qquad F_L = \frac{1}{4} (2 -A_0 \mp A_4)\, , \qquad F_R = \frac{1}{4} (2 -A_0 \pm A_4)\, ,
\end{equation}
for $W^\pm$, respectively. For electron-neutrino invariant mass far from the $W$ mass, we take Eq.\ \eqref{F0LR} as the definition of $F_{0,L,R}$.

In Fig.\ \ref{Fig:W2} we show the values of $F_0$ and $F_L  - F_R = -A_4/2$, for $m_W > m_{W\, \rm cut}$, where $m_W$ denotes the invariant mass of the lepton and the neutrino.
The error bands include scale and PDF variations. 
For illustration sake, we chose larger coefficients compared to Figs.\  \ref{Fig:mtW} and \ref{Fig:mZ}, but we stress that even for values closer to the limits discussed in Section \ref{bounds}
the effect on $F_{0,L,R}$ is significant.
In the SM, for small values of $p_T^W$, $A_0 \sim 0$ and $A_4$ is negative, reflecting the fact that  the $W$ boson is mostly produced with left-handed polarization \cite{Bern:2011ie}.
The values of $A_0$ and $A_4$ are not modified by the operators $C_{LQ, u}$ and $C_{LQ, d}$, which have the same chiral structure as the SM, and couple left-handed quark to left-handed lepton fields.
On the other hand, the dipole operators, and the scalar and tensor four-fermion operators have a different chiral structure.
In the case of the dipoles, the interaction of the $W$ to the quarks flips chirality. As a result the $W$ boson is mostly produced with longitudinal polarization, and,
if the dipole operators dominate the cross section, $F_0$ approaches $1$,  while $A_4$ should go to zero. 
This can explicitly be seen from the tree-level amplitude, that goes as $\sin^2\theta^*$. Indeed Fig.\ \ref{Fig:W2} shows that, in the presence of a non-vanishing $\Gamma_W^u$, 
as the $W$ invariant mass grows and the contribution of $\Gamma_W^u$ becomes more important, 
$F_0$ increases and $F_L - F_R$ decreases. Since, for $\Gamma_W^u = 0.05$, the total  cross section is  at most $50\%$ larger than in the SM, the polarization fractions shown in Fig.\ \ref{Fig:W2} do not reach the values expected when the dipole couplings dominate.
The cross section induced by a scalar interaction, such as  $C^{(1)}_{LeQu}$ and $C^{}_{LedQ}$, is, at LO, independent of $\cos\theta^*$. In this case, when the cross section is dominated by the scalar contribution, $F_L - F_R$ goes to zero and $F_0 \sim 1/3$, which is in agreement with Fig.\ \ref{Fig:W2}. 
Finally, for tensor interactions, the tree-level cross section goes as $\cos^2\theta^*$, suggesting that $F_0 \sim -1$ and $A_4 \sim 0$, which is once again confirmed by Fig.\ \ref{Fig:W2}.

While measurement of angular distributions in the $W$ rest frame are experimentally challenging, because of the incomplete  knowledge of the neutrino momentum, 
we hope that the discriminating power shown in Fig.\ \ref{Fig:W2}  motivates the  study of variables correlated with $\cos\theta^*$ and $m_W$ \cite{ATLAS:2012au},
which might reveal similar information.

\begin{figure}
\center
\includegraphics[width=0.475\textwidth]{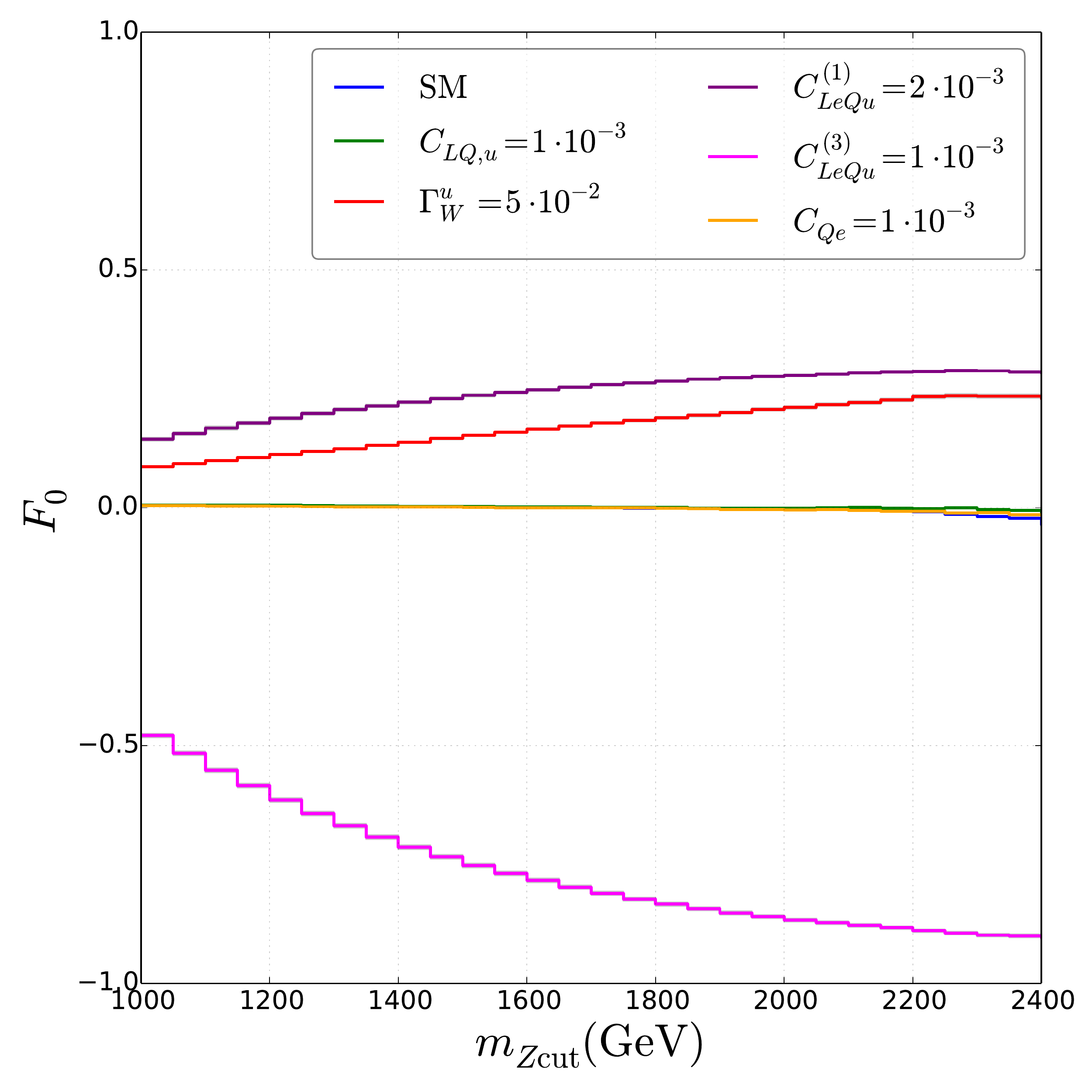}
\includegraphics[width=0.475\textwidth]{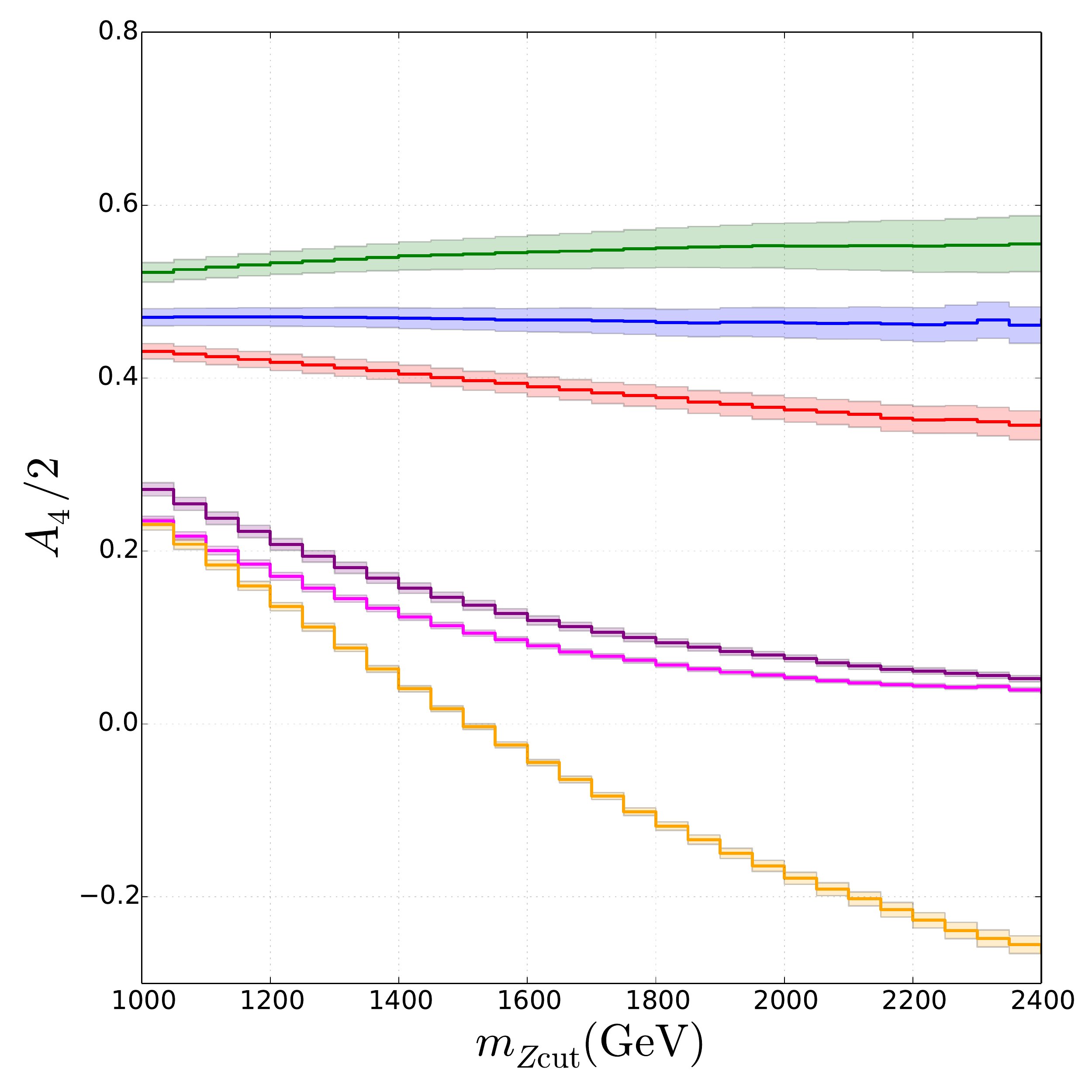}
\caption{Longitudinal polarization $F_0$ and angular coefficient $A_4$ in $pp\to e^+ e^-$ for $m_Z > m_{Z\, \rm cut}$, where $m_Z$ denotes the dilepton invariant mass, at $\sqrt{S} = 13$ TeV.
The shaded regions indicate the uncertainties from PDF and scale variations.}\label{Fig:Z2}
\end{figure}

For $Z$ production, we work in the Collins-Soper frame \cite{Collins:1977iv} and show $F_0$ and $A_4$ as a function of a cut on the leptons invariant mass in Fig.\ \ref{Fig:Z2}. 
In the SM $F_0$ is very close to zero, and does not significantly depend on the $Z$ invariant mass. $A_4$ is also close to zero at the $Z$ peak, 
and its dependence on $m_Z$ is determined by  $Z$/$\gamma^*$ interference. As the dilepton invariant mass grows, the SM cross section becomes dominated by the coupling to the $SU(2)_L$ gauge boson, resulting in a larger $A_4$. 
$C_{LQ, u}$ and $C_{LQ, d}$ do not change this picture, 
while new dipole, scalar, and tensor operators would dramatically change the value of $F_0$ and $A_4$, in the same way as in $W$ production.  
The operator $C_{Qe}$, which induces new couplings of left-handed quarks to right-handed leptons, 
does not change the longitudinal polarization, but, when it becomes dominant with respect to the SM, causes $A_4$ to flip sign.
Additional information can be extracted from the remaining angular coefficients, which encode the dependence on the azimuthal angle $\phi^*$.
This brief discussion shows that angular distributions can play a very important role in understanding the origin of new physics, were a deviation from the SM to be observed.

Finally, we note that it is straightforward to include the contributions of gauge-invariant dimension-six operators involving light sterile neutrinos $\nu_R$ \cite{Cirigliano:2012ab},
or  lepton-number violating (LNV) dimension-seven operators \cite{deGouvea:2007qla,Lehman:2014jma,Cirigliano:2017djv}.
For example, in the presence of $\nu_R$ it is possible to construct two more scalar ($\mathcal O_{qu\nu}$ and $\mathcal O^\prime_{lq}$ in the notation of Ref.\ \cite{Cirigliano:2012ab})  
and one more tensor operator  ($\mathcal O_{lq}^\prime$).  It is easy to see that, for charged-current processes,   
$\mathcal O_{qu\nu}$, $\mathcal O^\prime_{lq}$ and $\mathcal O_{lq}^\prime$ give rise to corrections to the squared amplitude that are identical to those of
$\mathcal O_{LedQ}$, $\mathcal O^{(1)}_{LeQu}$ and $\mathcal O^{(3)}_{LeQu}$, respectively.  Thus, were the  cross section of Fig.\ \ref{Fig:mtW} and the angular distributions  
of Fig.\ \ref{Fig:W2}  to show evidence of scalar or tensor operators, the effect can be attributed to non-standard operators with either left- or right-handed neutrinos (or even to LNV charged-current scalar and tensor operators).
In the first case, however, one would expect to see a correlated deviation in $p p \rightarrow \ell^+ \ell^-$. On the other hand, if the scalar or tensor operators involve $\nu_R$, or are LNV,  $SU(2)_L$ gauge-invariance relates the corrections to $p p \rightarrow \ell \nu_\ell$ to $p p \rightarrow \nu \bar{\nu}$, while $p p \rightarrow \ell^+ \ell^-$ is unaffected.


\section{Associated production of a Higgs boson and a $W$ or $Z$ boson}\label{HVprod}

\begin{table}
\begin{tabular}{||c|c|c|c|| c | c |c | c||}
\hline
 \multicolumn{4}{||c||}{Single coupling}  & \multicolumn{4}{c||}{Marginalized}\\
\hline 
$C_{\varphi W}$  & [-0.20,0.02] & $C_{\varphi \tilde W}$    & [-0.09,0.09] & $C_{\varphi W}$   & [-0.19,0.03] & $C_{\varphi \tilde W}$  & [-0.15,0.15]  \\
$C_{\varphi B}$  & [-0.28,0.15] & $C_{\varphi \tilde B}$    & [-0.26,0.26] & $C_{\varphi B}$   & [-0.06,0.01] & $C_{\varphi \tilde B}$  & [-0.05,0.05]  \\
$C_{\varphi WB}$ & [-0.42,0.07] & $C_{\varphi \tilde W B}$  & [-0.24,0.24] & $C_{\varphi W B}$ & [-0.22,0.03] & $C_{\varphi \tilde WB}$ & [-0.18,0.18]  \\
 $\Gamma_W^u$     &[-0.05,0.05] & $\Gamma_W^d$ 	        & [-0.06,0.06] & $\Gamma_W^u$      & [-0.13,0.13] & $\Gamma_W^d$ 	           & [-0.13,0.13]\\
$c^{}_{Q\varphi, U}$ & [-0.07,0.02] & $c^{}_{Q\varphi, D}$  & [-0.02,0.10] & $c^{}_{Q\varphi,U}$ & [-0.10,0.05] & $c^{}_{Q\varphi, D}$  & [-0.06,0.13] \\
$c^{}_{U\varphi}$ & [-0.03,0.05] & $c^{}_{D\varphi}$ 	& [-0.06,0.05] & $c^{}_{U\varphi}$ & [-0.06,0.09] & $c^{}_{D\varphi}$  & [-0.11,0.09] \\
$Y^\prime_u$             & [-0.04,0.04] &  		$Y^\prime_d$         & [-0.04,0.04] & $Y^\prime_u$ 	      & [-0.08,0.08] &$Y^\prime_d$  & [-0.08,0.08] \\
$\xi$             & [-0.04,0.04] &  		        & & $\xi$ 	      & [-0.06,0.06] & & \\
\hline
\end{tabular}
\caption{90\% CL level bounds on the coefficients of effective operators that contribute to $WH$ and $ZH$ production. For operators involving quarks, we turned on couplings to the $u$ and $d$ quark. 
The table on the left shows the bounds obtained assuming that a single operator is turned on at a scale $\Lambda =1$ TeV. On the right, marginalized bounds, 
from $WH$ and $ZH$ production and Higgs boson decays to $\gamma\gamma$, $\gamma Z$, $W W^*$ and $Z Z^*$.
}\label{HVbounds}
\end{table}

Associated production of a Higgs boson and a $W$  or $Z$  boson is the third
most important Higgs boson production mechanism at the LHC.  
Theoretical predictions for the SM background are available at NNLO in QCD~\cite{Brein:2003wg,Ferrera:2011bk,Ferrera:2013yga,Campbell:2016jau} and are matched to the parton shower up to the NNLO+PS level~\cite{Astill:2016hpa}.
Contributions from sets of SM-EFT operators, including NLO QCD corrections, have been studied in Refs.\ \cite{Maltoni:2013sma,Demartin:2014fia,Mimasu:2015nqa,Degrande:2016dqg,Greljo:2017spw}.
Refs.\ \cite{Maltoni:2013sma,Demartin:2014fia,Mimasu:2015nqa,Degrande:2016dqg} focused on couplings to the gauge bosons, such as $C_{\varphi W}$, $C_{\varphi B}$ and $C_{\varphi W B}$. Ref.\ \cite{Greljo:2017spw}
included also the quark operators $c^{}_{Q\varphi, U}$, $c^{}_{Q\varphi, D}$, $c_{U\varphi}$ and $c_{D\varphi}$, with the well-motivated flavor assumptions of universal couplings to the first two generations.
Compared to Ref.\ \cite{Greljo:2017spw}, the SM-EFT operators of Table \ref{TabB} include the right-handed current operator $\xi$, the Yukawa couplings $Y^\prime_{u,d}$ and the dipole operators $\Gamma^{u,d}_{W,\gamma}$.
In addition,  we allow  $c^{}_{Q\varphi, U}$, $c^{}_{Q\varphi, D}$, $c_{U\varphi}$ and $c_{D\varphi}$
to have quite general flavor structures, the only restriction being that the off-diagonal elements of these matrices are required to be real. For all the operators in Table \ref{TabB}  we include NLO QCD corrections, and interface with the parton shower, building upon the original NLO+PS \POWHEGBOX{} code in Ref.~\cite{Luisoni:2013kna}.

$WH$ and $ZH$ are very sensitive to the modifications of the $W$ and $Z$ boson couplings to quarks, written in a gauge invariant way in Eq.\ \eqref{eq:vertex}.
These operators require two scalar fields, which induce local quark-Higgs-gauge boson interactions that lead to a significant enhancement of the cross section.
Other operators that enter these processes are the dipole operators, since once
again $SU(2)_L$-invariance forces the presence of the Higgs field, and operators of the form $\varphi^{\dagger} \varphi X^{\mu \nu} X_{\mu \nu}$ 
and $\varphi^{\dagger} \varphi X^{\mu \nu} \tilde X_{\mu \nu}$.
The contributions of these three classes of operators to the cross section are
greatly enhanced at large $p_T$ of the Higgs boson or when the Higgs-weak boson invariant mass is large.
$WH$ and $ZH$ production also receive tree-level corrections from the non-standard Yukawa couplings $Y_{u,d}^\prime$. Even though these couplings are effectively dimension-four,
they induce a correction to the cross section that is not proportional to the mass of the vector boson, and thus is also enhanced by $p^2_T/m^2_{W,Z}$ at large $p_T$.

We can appreciate the sensitivity of  $WH$ and $ZH$ production to non-standard quark and gauge boson couplings by considering the $WH$ and $ZH$ signal strengths, defined as 
\begin{equation}
\mu_{WH} = \frac{\sigma_{W^+ H } + \sigma_{W^- H}}{ \sigma^{SM}_{W^+ H } + \sigma^{SM}_{W^- H}}, \qquad 
\mu_{ZH} = \frac{\sigma_{Z H } }{ \sigma^{SM}_{Z H } }.
\end{equation}
We computed the signal strength at $\sqrt{S} = 8$ TeV using the \texttt{PDF4LHC15$\_$nlo$\_$30} PDF sets \cite{Butterworth:2015oua},
and extracted bounds using the combined  results from the ATLAS and CMS collaborations \cite{Khachatryan:2016vau}
\begin{equation}
\mu_{WH}( 8 \textrm{TeV}) = 0.89^{+0.40}_{-0.38}, \qquad \mu_{ZH}( 8 \textrm{TeV}) = 0.79^{+0.38}_{-0.36}. 
\end{equation}
We estimated the scale uncertainties  on the signal strength by varying the renormalization and factorization scales $\mu_F$ and $\mu_R$ between $(m_V + m_H)/2$ and $ 2 (m_V + m_H)$,
where $m_V$ denotes the mass of the vector boson. 
The PDF  uncertainties were  estimated by evaluating the cross section using the 30 members of the \texttt{PDF4LHC15$\_$nlo$\_$30} PDF sets. 
For the operators we considered, the theoretical uncertainties are approximately $10\%$. 
These uncertainties were taken into account in the extraction of the constraints 
by following the Rfit approach \cite{Charles:2004jd}.

The left panel in Table \ref{HVbounds} shows the 90\% CL bounds on the coefficients on effective operators that contribute to $WH$ and $ZH$ production, under the assumption that only one operator is turned on at the new physics scale $\Lambda = 1 $ TeV. For simplicity we assumed that only the couplings to the $u$ and $d$ quark are affected. 
We see that the 8 TeV data, despite the large uncertainties, constrain non-standard couplings to light quarks at the $10\%$ level or better. 
The only exceptions are the  photon dipole operators $\Gamma^{u,d}_\gamma$, which give a small contribution to $ZH$ and are better constrained by NC Drell-Yan production.
Constraints on the couplings to gauge bosons $C_{\varphi W}$, $C_{\varphi B}$ and $C_{\varphi W B}$, and their CP-odd counterparts, $C_{\varphi \tilde{W}}$, $C_{\varphi \tilde{B}}$ and $C_{\varphi \tilde{W} B}$, 
are also weaker, ranging from $10\%$ to $40\%$.

If we simultaneously turn on all the operators, the limits on the quark operators $c^{}_{Q\varphi, U}$, $c^{}_{Q\varphi, D}$, $c_{U\varphi}$,  $c_{D\varphi}$, $\xi$, $Y_{u,d}^\prime$
and $\Gamma_W^{u,d}$ are weakened  by a factor of 2. The bounds on the gauge operators, in particular $C_{\varphi B}$ and $C_{\varphi WB}$, are more affected, becoming  
significantly weaker. On the other hand, these operators give tree level
corrections to the Higgs boson decays, both to $h \rightarrow WW^*$ and  $h \rightarrow ZZ^*$ as well as $h \rightarrow \gamma \gamma$ and $h \rightarrow \gamma Z$. Since the former two decays are generated at tree level in the SM while  the latter two   are loop-suppressed, the gauge operators are relatively more important for $h \rightarrow \gamma \gamma$ and $h \rightarrow \gamma Z$.
The expressions for the tree level corrections to $h \rightarrow \gamma \gamma$, $\gamma Z$, $W W^*$ and $Z Z^*$ induced by all the operators introduced in Section \ref{basis} are given in Appendix  \ref{AppDecay}.

In the right panel of Table \ref{HVbounds} we show the limits on the
coefficients of dimension-six operators from $WH$, $ZH$, and the Higgs boson branching ratios, 
obtained under the assumption that all couplings are present at the scale of new physics. 
We used  the combined Run-I ATLAS and CMS results for $\mu_{ h \rightarrow \gamma \gamma}$, $\mu_{ h \rightarrow W W^*}$
and $\mu_{ h \rightarrow Z Z^*}$ \cite{Khachatryan:2016vau}, and the ATLAS bound on $\mu_{h\rightarrow \gamma Z}$ \cite{Aaboud:2017uhw} 
\begin{equation}
\mu_{h \rightarrow \gamma \gamma}  = 1.14^{+0.19}_{-0.18} , \qquad \mu_{h \rightarrow W W^*} =  1.09^{+0.18}_{-0.16} \qquad \mu_{h \rightarrow Z Z^*}  =  1.29^{+0.26}_{-0.23}\, \qquad \mu_{h \rightarrow \gamma Z} < 6.6.  
\end{equation}
We see that  the Higgs boson branching ratios provide better constraints on $C_{\varphi B (\tilde B)}$ and $C_{\varphi WB (\tilde WB)}$, while they have no significant effect on the remaining operators, for which the $WH$ and $ZH$ cross sections are more sensitive observables. Additional constraints on the Higgs-gauge operators come from electroweak precision observables \cite{Alonso:2013hga},
while strong constraints on non-standard Yukawas, at the $\sim 1\%$ level, 
can be obtained from the total Higgs boson production cross section and from
the Higgs boson decays \cite{Perez:2015aoa,Chien:2015xha,Soreq:2016rae}.

\begin{figure}
\includegraphics[width=15cm]{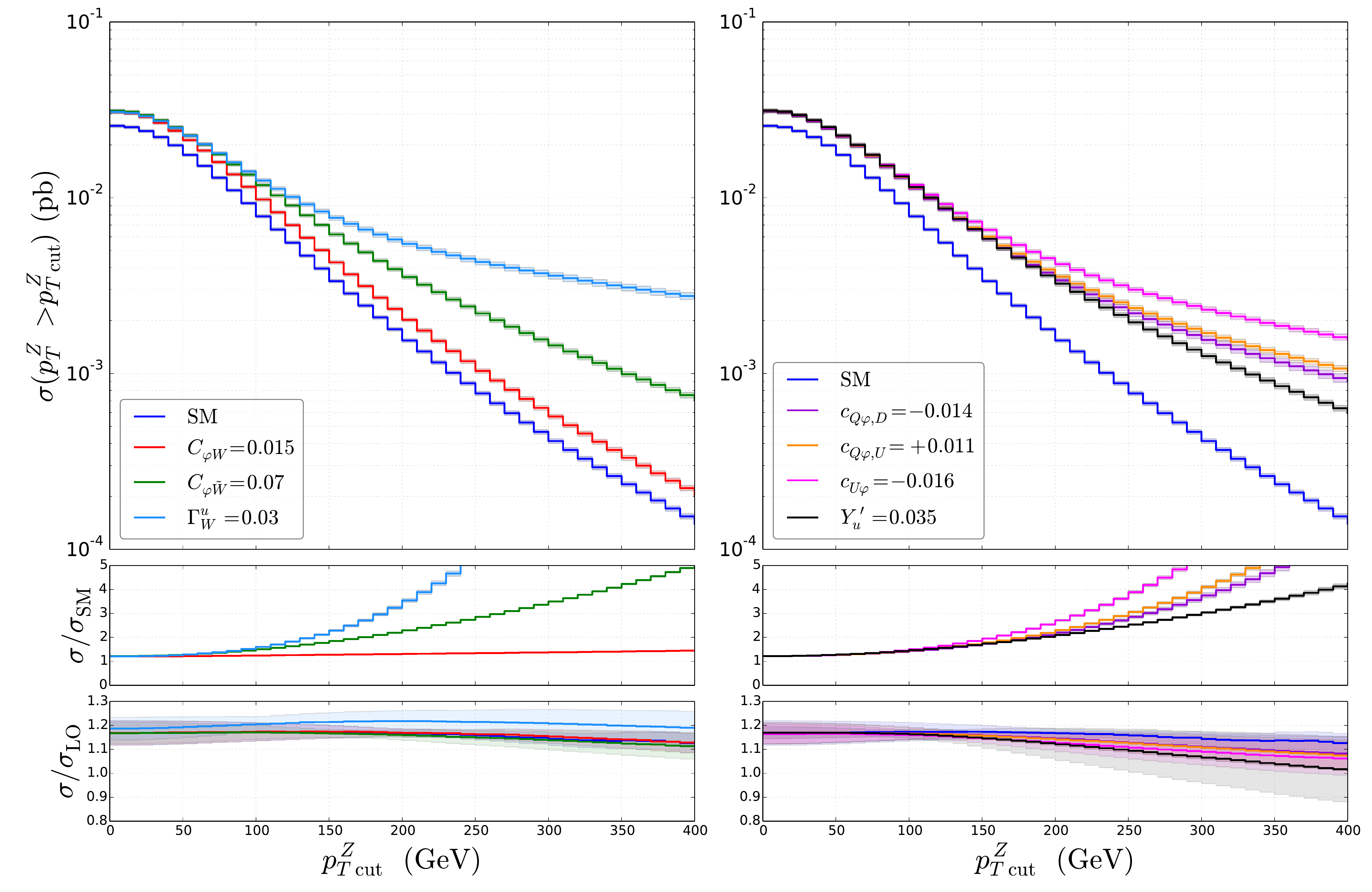}
\caption{Cumulative $ZH$ production cross section for $p^Z_{T} > p^Z_{T\, \textrm{cut}}$, at $\sqrt{S} = 13$ TeV.  }\label{Fig:Ass1}
\end{figure}

We have so far only made use of the total $WH$ and $ZH$ cross sections. More differential information, such as the $p_T$ spectrum of the Higgs boson or the distribution in invariant mass 
of the Higgs and vector boson, will allow to further tighten the constraints on the effective operators. In addition, in the presence of  deviations from the SM expectations, such information
would help  to disentangle the possible new physics mechanisms.
In Ref.~\cite{Alioli:2017ces} we  discussed in detail $WH$ production, and suggested that the angular distributions of the charged lepton in the $W$ rest frame can be used to disentangle the different dimension-six operators.
We further identified two angular coefficients, $A_3$
and $A_5$, which would pinpoint, respectively, the coupling of the $W$ to right-handed quarks of the first generation, $\xi_{ud}$, or the CP-odd operator $\varphi^{\dagger} \varphi W_{\mu \nu} \, \tilde{W}^{\mu \nu}$. 
The same discussion can be extended to $ZH$ production.

In Fig.\ \ref{Fig:Ass1} we show the cumulative $ZH$ cross section, as a function of a cut on the $Z$ transverse momentum, for $\sqrt{S} = 13$ TeV.
The blue line denotes the SM cross section, while we set the couplings of dimension-six operators so that the total cross section with no cut on $p_T^{Z}$ is 20\% larger than the SM.
The bands denote scale and PDF uncertainties, estimated using the \texttt{PDF4LHC15$\_$nlo$\_$30} PDF sets. 
For the operators we considered the theoretical uncertainties are around $5\%$-$10\%$, and roughly independent on the value of $p^{Z}_{T\, \rm cut}$.
The middle panel of Fig.\ \ref{Fig:Ass1} shows that for all dimension-six operators the contribution to the cross section increases at large $p_T$.  The increase is most marked for the dipole operators $\Gamma_W^{u,d}$. In this case, if one neglects small SM Yukawas, 
there is no interference with the SM, and the  contribution of the dipole operators  is dominated by diagrams in which the $Z$ and Higgs boson are created directly in the $q\bar q$ annihilation.
These diagrams are enhanced by $(s/v^2)^3$ with respect to the SM, where $\sqrt{s}$ is the partonic center of mass energy.
Similarly, the interference of $C_{\varphi \tilde W}$ with the SM does not contribute to the cumulative cross section. The correction is thus quadratic in $C_{\varphi \tilde W}$,
and, for large $s$, is enhanced by $(s/v^2)^2$ with respect to the SM.
In the case of $C_{\varphi W}$ and of the vertex corrections $c^{}_{Q\varphi, U}$, $c^{}_{Q\varphi, D}$, $c_{U\varphi}$ and $c^{}_{D\varphi}$,
there is competition between the interference and the quadratic pieces, 
which are enhanced by $s/v^2$ or $(s/v^2)^2$ compared to the SM, respectively. 
Finally, the non-standard Yukawa cross section is also enhanced by $p^2_T/m_{Z,W}^2$ at large $p_T$.
Data on differential distributions from the LHC Run II would therefore be extremely helpful in constraining the effective operators in Table \ref{HVbounds} at the few percent level.

The bottom panel of Fig.\ \ref{Fig:Ass1} shows the ratio of the NLO and LO cross sections, for the SM and dimension-six operators.
In most cases the NLO corrections are about 10\%-20\% across the values of $p^Z_{T\, \textrm{cut}}$ considered in Fig.\ \ref{Fig:Ass1}. The exceptions are non-standard Yukawa couplings, for which the NLO corrections  decrease at higher $p^Z_{T\, \textrm{cut}}$.

As in the case of $WH$ production, angular distributions of the leptons emitted in the decay of the $Z$ boson can help disentangle the effects of different dimension-six operators. 
We work in the $Z$-boson rest frame,  with the direction of the $z$-axis along the momentum of the $Z$ boson in the lab frame. $\theta^*$ is the polar angle of the electron in this frame.
The  $x$-axis is in the direction orthogonal to the Higgs boson and $Z$ momenta $\hat{x} \sim (\vec{p}_Z \times \vec{p}_H)$. In this frame, we define the azimuthal angle $\phi^*$ as the angle between the plane containing the $Z$ and the Higgs bosons, and the plane containing the $Z$ and the electron. That is
\begin{equation}
\cos\phi^* = \frac{(\vec p_Z \times \vec p_H) \cdot (\vec p_Z \times \vec p_{e^-})}{|\vec p_Z \times \vec p_H|\,  |\vec p_Z \times \vec p_{e^-}|} \, ,
\end{equation}
and we note that $\phi^*$ is invariant under boosts along the $Z$ momentum $\vec p_Z$. 
The angular distribution of the $Z$ boson in this frame is parameterized by 8 coefficients
\begin{eqnarray}
& & \frac{1}{\sigma} \frac{d\sigma}{d\cos\theta^*\, d\phi^*} = \frac{3}{16 \pi} \Big[  1 + \cos^2\theta^*  + \frac{A_0}{2} (1- 3 \cos^2 \theta^*) + A_1 \sin 2\theta^* \cos\phi^*  
+ \frac{A_2}{2} \sin^2\theta^* \cos 2\phi^*   \nn \\ &&  + A_3 \sin\theta^* \cos\phi^* + A_4 \cos\theta^* + A_5 \sin\theta^* \sin \phi^*  + A_6 \sin2\theta^* \sin \phi^*
+ A_7 \sin^2\theta^* \sin 2\phi^* \label{diffthetaphi}
\Big]\, .
\end{eqnarray}
The coefficients  $A_0$ and $A_4$ are related to the $Z$-boson helicity fractions \cite{Stirling:2012zt,Karlberg:2014qua,Astill:2016hpa}. 
\begin{equation}
F_0 = \frac{A_0}{2}\, , \qquad F_L - F_R = -\frac{A_4}{2}  \, \alpha .
\end{equation}
The difference between left- and right-handed polarization depends on $A_4$ and on the ratio $\alpha = (c_L^2 - c_R^2)/(c_L^2 + c_R^2)$, where $c_{L,R}$ are the couplings of the $Z$ boson to left- and right-handed leptons. 
The operators we study do not modify these couplings at tree level, and $\alpha$  is fixed by the Weinberg angle, $\sin\theta_W$.

\begin{figure}
\includegraphics[width=15cm]{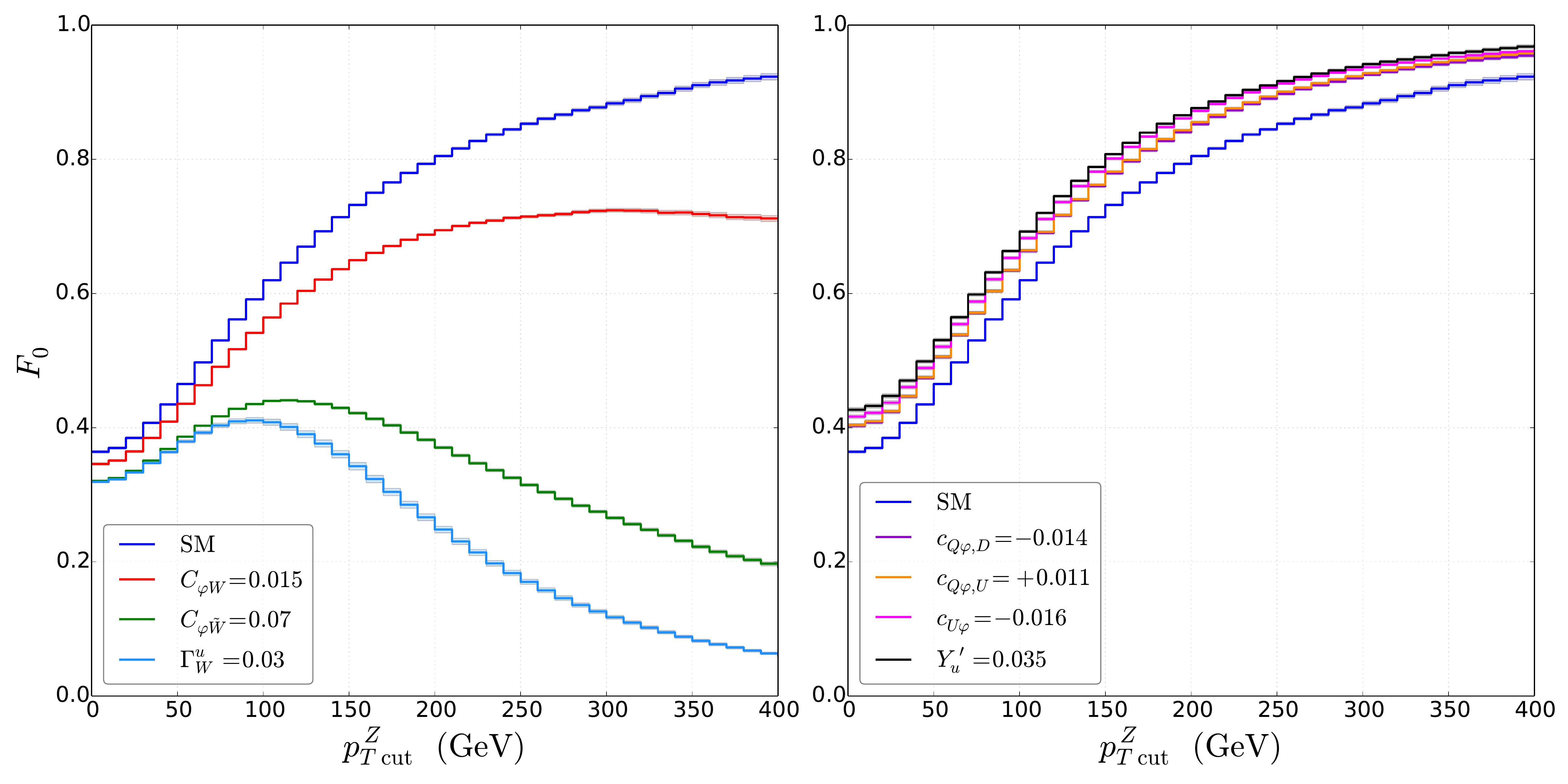}
\caption{Longitudinal polarization of the $Z$ boson in $ZH$ as a function of $p^{Z}_{T\, \textrm{cut}}$, at $\sqrt{S} = 13$ TeV.}\label{Fig:Ass2}
\end{figure}

In Fig.\ \ref{Fig:Ass2} we show $F_0$ as a function of a cut on the $Z$ transverse momentum. The figure shows that the $Z$ boson is produced with a high degree of longitudinal polarization in the SM, and if the SM is modified by the couplings $c^{}_{Q\varphi, U}$, $c^{}_{Q\varphi, D}$, $c_{U\varphi}$ ( and $c_{D\varphi}$, which is not shown in the figure).  These couplings have the same helicity structure as the SM $Z$ couplings,
and thus it is not surprising that they do not affect the polarization much. The situation is quite different for dipole operators, that prefer the $Z$ boson to have transverse polarization, pushing $F_0$ to zero at large $p_T$.
In the case of  $WH$ production, the operators $\Gamma_W^u$ and $\Gamma^d_W$ produce the $W$ boson in an almost complete left-handed and right-handed polarized state, respectively \cite{Alioli:2017ces}.
In the case of $ZH$ production, the difference between left- and right-polarization, captured by the coefficient $A_4$, is always small, so that both $\Gamma^u_W$ and $\Gamma^d_W$ produce a right-polarized $Z$ boson
approximately half of the times.
The operator  $C_{\varphi \, \tilde W}$ also prefers the $Z$ to be transversely polarized.  
The theoretical uncertainties from PDF and scale variations, shown by the bands in Fig.\ \ref{Fig:Ass2}, are small.

\begin{figure}
\center
\includegraphics[width=9cm]{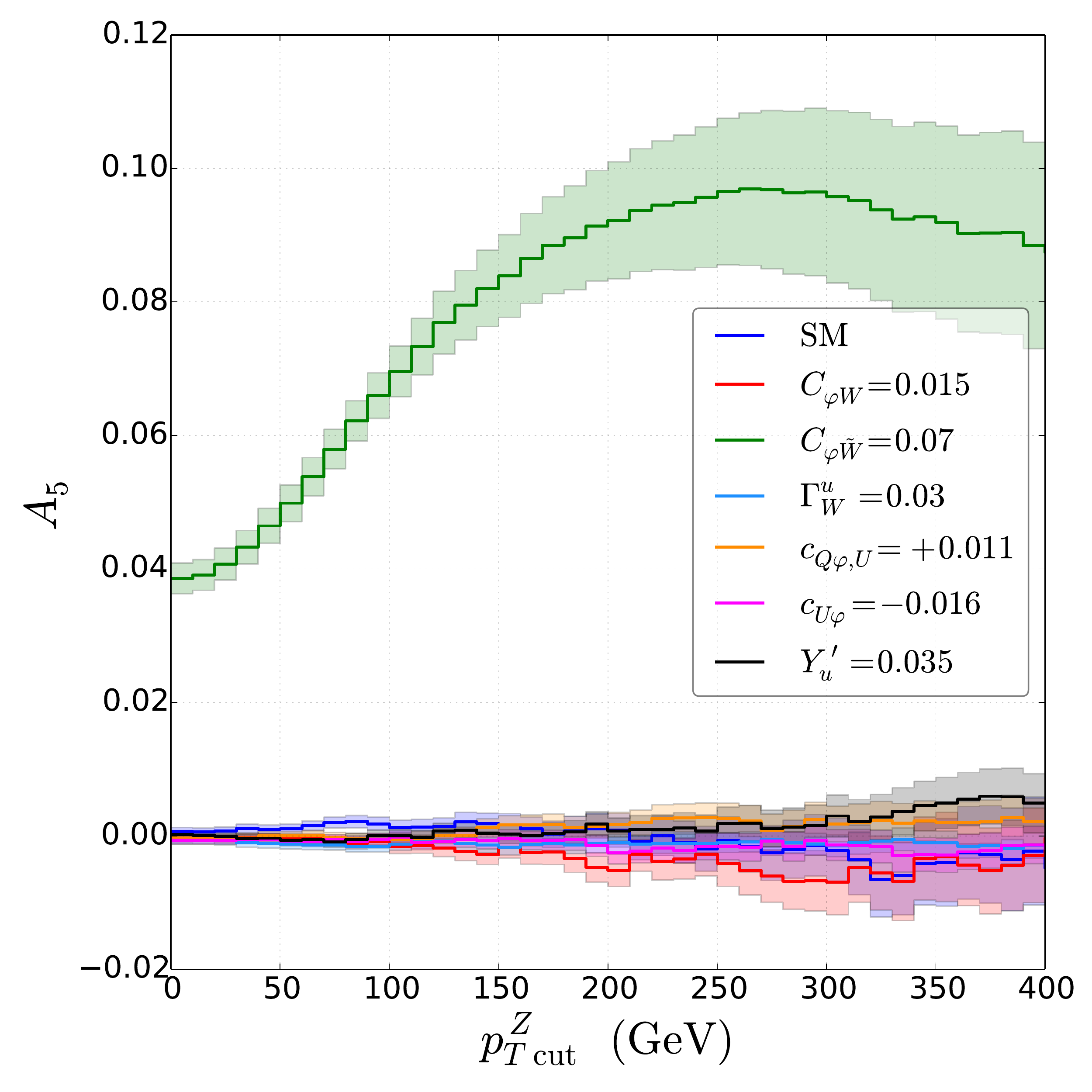}
\caption{Angular coefficient $A_5$ as a function of $p^{Z}_{T\, \textrm{cut}}$, at $\sqrt{S} = 13$ TeV. 
}\label{Fig:Ass3}
\end{figure}

The remaining coefficients in Eq.\ \eqref{diffthetaphi} also carry important information. For example, $A_5$ is sensitive to CP violation. $A_5$ vanishes in the SM, at NLO in QCD, but it receives 
a relatively large contribution from  the interference of the operator $C_{\varphi \tilde{W}}$ with the SM, as shown in Fig.\ \ref{Fig:Ass3}. The remaining dimension-six operators, at least with the 
assumptions of Sec.\ \ref{basis} on their flavor structure, do not contribute to $A_5$, thus making it a clean observables to pinpoint  $C_{\varphi \tilde{W}}$.
Notice however that $A_5$ is affected by a larger theoretical error, which is dominated by scale variations.


\section{Vector boson fusion}\label{VBF}

\begin{figure}
\includegraphics[width=\textwidth]{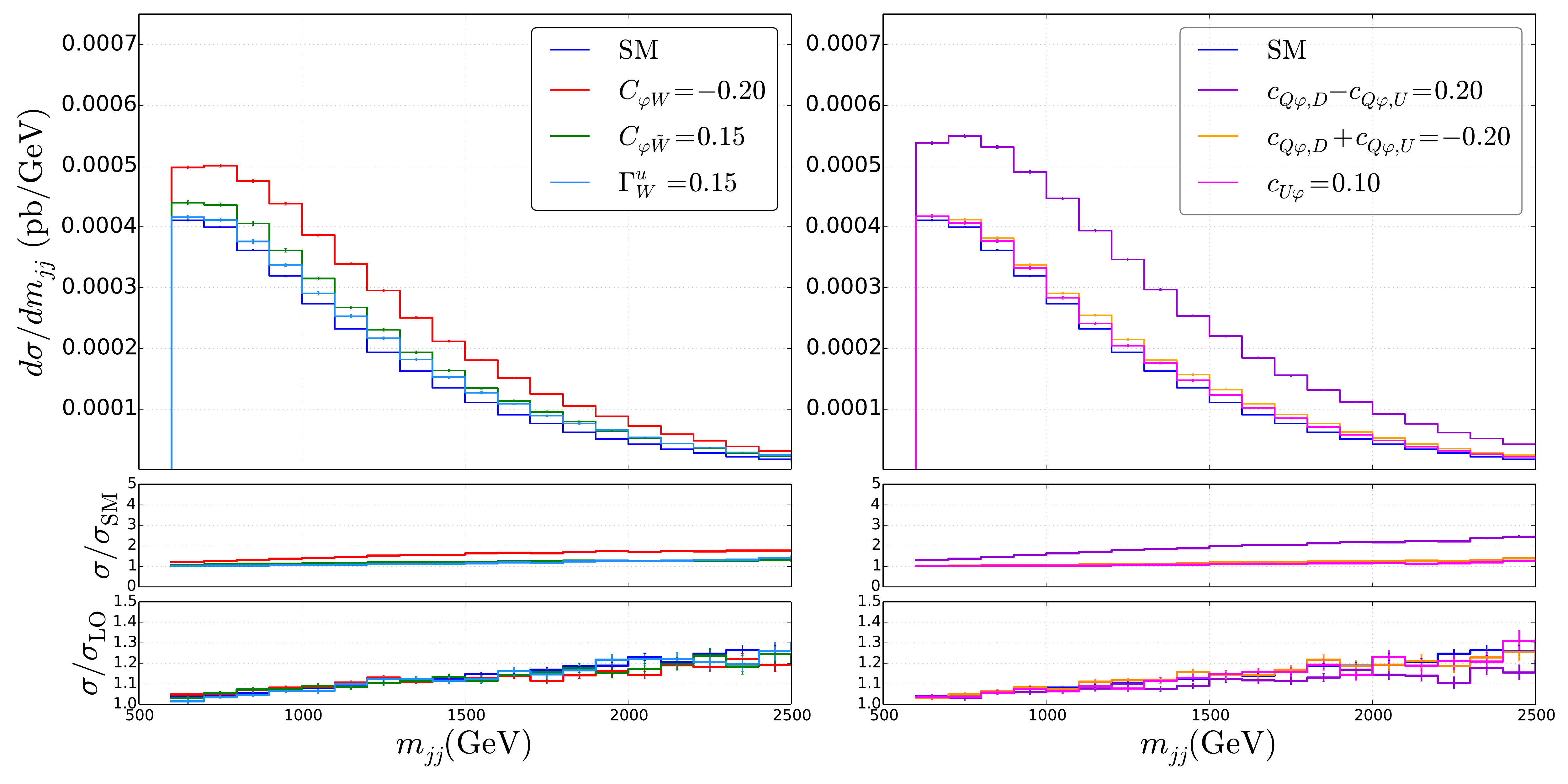}
\caption{Differential VBF cross section, as a function of the dijet invariant mass $m_{jj}$, at $\sqrt{S} = 8$ TeV, with the standard VBF cuts described in the text.}\label{FigVBF2}
\end{figure}

Finally we discuss the corrections to Higgs boson production via vector boson fusion.
The total cross section for Higgs boson production through VBF has been recently computed in the SM at N${}^3$LO in QCD~\cite{Dreyer:2016oyx}. Fully-differential distributions are available up to NNLO~\cite{Cacciari:2015jma} and the interface with parton showering is available at NLO+PS accuracy~\cite{Nason:2009ai}. 
For this study, we computed the NLO QCD corrections to both the SM and the dimension-six SM-EFT contributions to the VBF cross sections,
building upon the \POWHEG{} implementation presented in Ref.\ \cite{Nason:2009ai}. The contribution of SM-EFT operators, including NLO QCD corrections, has been considered in Ref.\ \cite{Maltoni:2013sma,Demartin:2014fia,Greljo:2017spw}.
Here we study the operators in Table \ref{TabB}. 
In addition to the flavor assumptions discussed in Sec.\ \ref{sec:flavorStructure}, in VBF we impose the further restriction of no tree-level flavor-changing neutral currents,
so that only the diagonal components of $\Gamma^{u,d}_{W,\gamma}$, $c_{Q\varphi, U}$, $c_{Q\varphi, D}$, $c_{U\varphi}$ and $c_{D\varphi}$ can be non-zero. 
This additional assumption arises naturally in well motivated flavor models, such as minimal flavor violation \cite{D'Ambrosio:2002ex}, where off-diagonal elements are suppressed
by small CKM elements or Yukawa couplings, and are irrelevant for collider observables. 
This assumption has the practical advantage of reducing the channels to be considered.

As in Ref.\ \cite{Nason:2009ai}, we neglect contributions from $VH$ production, with the vector boson  decaying hadronically, and,
for channels with two identical quarks in the final state, 
we neglect effects due to the interchange of identical quarks. 
These effects are suppressed in the experimentally interesting region with two widely separated jets of large invariant mass.
In the implementation of Ref.\ \cite{Nason:2009ai}, \POWHEG{} generates events with the CKM matrix set to the identity, thus reducing the number of Feynman diagrams and singular regions.
The final state quarks are then reweighted by the SM CKM matrix before they are showered by a Monte Carlo program.
One could use the same approach for dimension-six operators that do not couple to quarks, such as $C_{\varphi W}$. On the other hand, if the SM-EFT operators couples to quarks, such reweighting is not possible, and one needs to generate events with the proper flavor configuration.

Some of the operators described in Section \ref{basis}, such as $\Gamma_\gamma^{u,d}$ or $C_{\varphi B}$, induce contributions  
to VBF mediated by the exchange of a photon. These contributions can generate collinear divergences, which 
are cut-off by the $p_T$ and rapidity cuts in the experimentally relevant region, but can make the generation of events extremely inefficient. 
In order to avoid these singular regions, if the coefficients of operators involving photons are turned on, the generation of events should be performed setting the  $\texttt{bornsuppfact}$ flag to an appropriate value in the \texttt{powheg.input} file.

In Fig.\ \ref{FigVBF2} we plot the differential cross section with respect to the dijet invariant mass, at $\sqrt{S} = 8$ TeV. 
We applied the standard VBF cuts, requiring  the invariant mass of the two jets  to be $m_{j j} > 600$ GeV,   the rapidity separation $|y_{j_1} - y_{j_2}|  >4.2$, and that all jets have $p_{Tj} > 20$ GeV and $|y_j| < 5$. 
We computed the cross section at the renormalization scale $\mu = 2 m_Z$, and used the first member of \texttt{PDF4LHC15$\_$nlo$\_$30} PDF set.
The error bars in Fig.\ \ref{FigVBF2} only reflect the statistical error of the Monte Carlo integration.
The blue curve denotes the SM cross section, while the remaining curves depict the effect of dimension-six operators, turned on one at a time. 
The coefficients are chosen to be close to the values excluded by $VH$ production, shown in Table \ref{HVbounds}.
For these values the correction to the VBF signal strength goes from 10-15\% in the case of $C_{\varphi \tilde W}$, $\Gamma_W^u$,  $c^{}_{Q\varphi, U} + c^{}_{Q\varphi,D}$ and $c^{}_{U\varphi}$,
to $40\%$ in the case of $C_{\varphi W}$ and $60\%$ in the case of $c^{}_{Q\varphi, D} - c^{}_{Q\varphi, U}$. 
The  VBF signal strength measured  at the LHC Run I  \cite{Khachatryan:2016vau}
allows for such deviations, so that, with the exception of  $c^{(3)}_{Q\varphi}$, including VBF 
does not improve the constraints discussed in Section \ref{HVprod}.

The lower sensitivity of VBF compared to $ZH$ and $WH$ production can be appreciated from the bottom panel of Fig.\ \ref{FigVBF2}, which shows that  
the VBF cross section induced by SM-EFT operators grows slowly as the dijet invariant mass increases.  One advantage of VBF is the possibility to flavor tag the jets in the final states,  
providing an additional handle on the flavor structure of SM-EFT operators.  

Finally, the bottom panel of Fig.\ \ref{FigVBF2} shows the ratio of the NLO and LO predictions. Also in this case NLO corrections range between 10\%  and 30\%.

\section{Conclusion}\label{conclusion}

The SM-EFT is a powerful tool to characterize deviations from the SM, under
the assumption that new physics arises at scales heavier than directly
accessible at the LHC.  In this paper, we have discussed the corrections of
SM-EFT dimension-six operators to Drell-Yan, associated production of the
Higgs boson and a weak boson, and Higgs boson production via vector boson
fusion.  We have included NLO QCD corrections, and interfaced to the parton
shower using the \POWHEG{} method. The resulting \POWHEGBOXVTWO{} computer
code will be made publicly available on the \texttt{powhegbox.mib.infn.it}
webpage.  As possible applications, we have discussed the constraints on
SM-EFT operators from the NC and CC Drell-Yan cross sections at high dilepton
or transverse mass, and the $ZH$, $WH$ and VBF signal strengths.  We find that
high-mass NC and CC Drell-Yan put strong constraints on new vector, axial,
scalar or tensor semileptonic contact operators, which are now competitive
with the low-energy constraints from pion and kaon decays and nuclear $\beta$
decay. $ZH$, $WH$ and VBF are on the other hand sensitive to modifications of
the Higgs couplings and to new couplings of the Higgs and vector bosons to
quarks. The Higgs boson signal strengths at the LHC Run I, notwithstanding
their relatively large uncertainties, already suggest that the scale of these new
interactions has to be larger than $1$ TeV. The constraints can be
significantly improved by measurements of the signal strength and, more
importantly, of differential distributions as the Higgs boson $p_T$ spectrum
at the LHC Run II.  We have pointed out that differential and angular
distributions will play an important role in removing the degeneracy between
different SM-EFT operators, thus providing clues on the form of physics beyond the SM.

\section*{Acknowledgements}
We thank Admir Greljo and Fabio Maltoni for comments on the manuscript. 
SA acknowledges support by the ERC Starting Grant REINVENT-714788 and by 
Fondazione Cariplo and Regione Lombardia, grant 2017-2070.
EM and WD  acknowledge support by the US DOE Office of Nuclear Physics and by the LDRD program at Los Alamos National Laboratory.
WD acknowledges  support by the Dutch Organization for Scientific Research (NWO) 
through a RUBICON.

\appendix

\section{Formulae for the Higgs boson decay}\label{AppDecay}

The operators in Table \ref{TabB} that contribute to $WH$, $ZH$ and VBF also
give tree level corrections to the Higgs boson decays to $\gamma \gamma$, $\gamma Z$, $W W^*$ and $Z Z^*$.
In this appendix, we give the relevant formulae.

\subsection{$h \rightarrow \gamma \gamma$ and $h \rightarrow \gamma Z$}

In the SM the decays $h \rightarrow \gamma\gamma$ and $h \rightarrow \gamma Z$
are mediated by loops involving Higgs boson couplings to the top quark and to the weak bosons. 
In the SM-EFT the decay widths are modified, at tree level, by the operators $C_{\varphi W}$, $C_{\varphi B}$ and $C_{\varphi W B}$,
and by their CP-odd counterparts, $C_{\varphi \tilde W}$, $C_{\varphi \tilde B}$ and $C_{\varphi \tilde W B}$. We find  

\begin{eqnarray}
\frac{\Gamma_{h \rightarrow \gamma \gamma} }{ \Gamma^{SM}_{h \rightarrow \gamma \gamma} } &=&  \frac{\left( N_c Q_t^2  - \frac{21}{4} A(\tau_W) + \frac{24 \pi^2}{e^2} C_{\varphi \gamma} \right)^2 + \left(\frac{24 \pi^2}{e^2} C_{\varphi \tilde{\gamma}}\right)^2 }{\left( N_c Q_t^2  - \frac{21}{4} A(\tau_W) \right)^2} 
\end{eqnarray}

\begin{eqnarray}
\frac{\Gamma_{h \rightarrow \gamma Z} }{ \Gamma^{SM}_{h \rightarrow \gamma Z} } &=&  \frac{\left( A_t^H(\tau_t,\lambda_t) + A_W^H(\tau_w,\lambda_w) - \frac{16 \pi^2}{e g} C_{\varphi \gamma Z} \right)^2 + \left(\frac{16 \pi^2}{e g} C_{\varphi \tilde{\gamma} Z}\right)^2 }{\left(A_t^H(x_t,y_t) + A_W^H(x_w,y_w) \right)^2}
\end{eqnarray}
where 
\begin{eqnarray}
C_{\varphi \gamma  \, (\tilde \gamma)} &=& s_w^2 C_{\varphi W \, (\tilde W)}  + c_w^2 C_{\varphi B (\tilde B)} - c_w s_w C_{\varphi W B \, (\tilde W B)}\nn \\
C_{\varphi \gamma Z \, (\tilde \gamma Z)} &=& 2 c_w s_w  ( C_{\varphi W \, (\tilde W)}  -  C_{\varphi B \, (\tilde B})) - ( c_w^2 - s^2_w ) C_{\varphi W B \, ( \tilde W B)}. 
\end{eqnarray}
$N_c = 3$ denotes the number of colors, $Q_t = 2/3$ the top charge, and we are neglecting contributions from quarks lighter than the top.
The variables $\tau$ and $\lambda$ are $\tau_t  = 4 m_t^2/m_H^2$, $\tau_w = 4 m_W^2/m_H^2$, $\lambda_t = 4 m_t^2/m_Z^2$ and $\lambda_w = 4 m_W^2/m_Z^2$.  
The loop functions $A_t^H$ and $A_W^H$ are given in Ref.\ \cite{Spira:1997dg}:
\begin{eqnarray}
A_t^H(\tau,\lambda) &=& 2 N_c Q_t \frac{1}{c_w} \left(\frac{1}{2} - 2 Q_t s_w^2\right) \left(I_1 (\tau,\lambda) - I_2 (\tau,\lambda)\right) \nn \\
A_w^H(\tau,\lambda) &=& c_w \left\{  4 \left(3 - \frac{s_w^2}{c_w^2}\right) I_2(\tau,\lambda) + \left[ \left(1 + \frac{2}{\tau} \right) \frac{s_w^2}{c_w^2} - \left(5 + \frac{2}{\tau}\right)\right] \, I_1(\tau,\lambda)  \right\},
\end{eqnarray}
with 
\begin{eqnarray}
I_1(\tau,\lambda) &=& \frac{\tau \lambda}{2 (\tau - \lambda)} + \frac{\tau^2 \lambda^2}{2 (\tau - \lambda)^2} (f(\tau) - f(\lambda) ) + \frac{\tau^2 \lambda}{(\tau - \lambda)^2} (g(\tau)  - g(\lambda)) \nn \\
I_2(\tau,\lambda) &=& - \frac{\tau \lambda}{2 (\tau - \lambda)} (f(\tau) - f(\lambda) ). 
\end{eqnarray}
Finally, $f$ and $g$ are, for $\tau > 1$
\begin{eqnarray}
f(\tau) = \arcsin^2 \left(\frac{1}{\sqrt{\tau}}\right) \qquad g(\tau) = \sqrt{\tau -1} \, \arcsin \left(\frac{1}{\sqrt{\tau}}\right).
\end{eqnarray}

\subsection{$h \rightarrow W W^*$ and $h \rightarrow Z Z^*$}
The decay rates for $h \rightarrow W W^*$ and $h \rightarrow Z Z^*$ are induced  at tree level in the SM, and are modified by all the operators in Table \ref{TabB} that affect $ZH$ and $WH$ production.
Including terms quadratic in the coefficient of effective operators, we find 

\begin{eqnarray}
& & \Gamma(h \rightarrow W W^*) = \frac{3 m_H m_W^4}{32 \pi^3 v^4}  \times \nn \\
& & \Bigg\{   \Bigg( 1 +  \sum_{i,j}   |V_{ij}|^2 \Bigg) \left( R(x) + C_{\varphi W} R_{W}(x)  + C_{\varphi W}^2 R_{WW}(x) + C_{\varphi \tilde{W}}^2 R_{W\tilde{W}}(x)\right)   \nn \\ 
&&-\sum_{ij}\frac{1}{2}{\rm Re}\left[V_{ij}^*\left(Vc_{Q\vp,D}-c_{Q\vp,U}V\right)_{ij}\right]\, R_{Q\varphi}(x) 
 \nn \\ 
&&+ \sum_{ij}    \left(  |\xi_{ij}|^2  + \frac{1}{4} \left| \left( Vc_{Q\vp,D}   -  c_{Q\vp, U}V \right)_{ij} \right|^2  \right) R_\xi(x)  
  \nn\\
&&+g\sq \sum_{ij}\left(\big| \big(V^\dagger \Gamma^u_W\big)_{ji}\big|\sq+\big|
\big(V\Gamma^d_W\big)_{ij}\big|\sq\right)R_\Gamma(x)
\Bigg\} \label{WWstar}
\end{eqnarray}
\begin{eqnarray}
& &\Gamma(h \rightarrow Z Z^*) = \frac{3 m_H m_Z^4}{32 \pi^3 v^4} \times \nn \\  & & \Bigg\{   \left( g_L^{(l)\, 2} + g^{(l)\, 2}_R + g_L^{(\nu)\, 2} +  \sum_{q  }   (g_L^{(q) \, 2}  + g_R^{(q)\, 2}) \right) \left( R(x) + C_{\varphi Z} R_{W}(x)   
 + C_{\varphi Z}^2 R_{WW}(x)\right.  \nn \\ &&  \left.  + C_{\varphi \tilde{Z}}^2 R_{W\tilde{W}}(x)\right)   \nn \\ 
 & &   + \frac{1}{2} \left(  \sum_{i} g_L^{(i)}  \left( c_{Q\varphi,U} \right)_{ii}
 + \sum_{j} g_L^{(j)}   \left( c_{Q\varphi,D} \right)_{jj} 
 + \sum_{i} g_R^{(i)} \, \left(c^{}_{U\varphi}\right)_{ii} + \sum_{j} g_R^{(j)} \, \left(c^{}_{D\varphi}\right)_{jj}  \right) \nn\\  & &  \times \, R_{Q\varphi}(x) 
 \nonumber \\ & &   +   \frac{1}{4}  \left( \sum_{ik} \left(\big|\left( c_{Q\varphi,U} \right)_{ik}\big|^2 +\big|\left( c_{U\varphi} \right)_{ik}\big|\sq\right) 
+  \sum_{jl}\big| \left(\left( c_{Q\varphi,D} \right)_{jl}\big|^2 + \big|\left(c_{D\varphi} \right)_{jl}\big|^2\right) \right) R_\xi(x)  \nn\\
&&+\frac{g\sq}{c_w\sq}\left( \sum_{ik}\big|\left( \Gamma^u_W-s_w\sq \Gamma_\gamma^u\right)_{ik}\big|\sq+\sum_{jl}\big|\left( \Gamma^d_W+s_w\sq \Gamma_\gamma^d\right)_{jl}\big|\sq\right)R_\Gamma(x)
 \Bigg\}\ ,    
\label{ZZstar}
\end{eqnarray}
where $x = m_W^2/m_H^2$ in Eq.\ \eqref{WWstar} and  $x = m_Z^2/m_H^2$ in Eq.\ \eqref{ZZstar}, and $V$ denotes the CKM matrix. The sums over $i,k$ and $j,l$ extend over all light quarks,  $i,k \in \{u,c\}$ and $j,l \in \{d,s,b\}$. 
$g^{(f)}_L$ and $g^{(f)}_R$ are the SM couplings of the $Z$ boson to left- and right-handed fermions 
\begin{equation}
g^{(f)}_L = T_3^{f} - Q_f s_w^2, \qquad g^{(f)}_R = - Q_f s_w^2, 
\end{equation}
and $C_{\varphi Z}$ and $C_{\varphi \tilde{Z}}$ are combinations of couplings
\begin{equation}
C_{\varphi Z\, (\tilde Z)} = c_w^2 \, C_{\varphi W\, (\tilde W)} + s_w^2 C_{\varphi B\, (\tilde{B})}  + c_w s_w C_{\varphi W B\, (\tilde W B)}.
\end{equation}

The functions are defined as  

\begin{eqnarray}
R(x) &=& -\frac{1 - x}{6 x} (2 - 13 x + 47 x^2) - \frac{1}{2} (1 - 6 x + 4 x^2) \log x  + \frac{ (1 - 8 x + 20 x^2)}{\sqrt{4 x - 1}} \arccos\left( \frac{3 x -1}{2 x^{3/2}} \right), \nn \\
R_{Q\varphi}(x) & = &  -\frac{1 - x}{9 x} (17 - 100 x + 179 x^2)  + \frac{1}{3 x} (1 - 9 x + 54 x^2 - 12 x^3) \log x \nn \\
& & - \frac{ 6 - 66 x + 312 x^2 - 648 x^3}{9 x \sqrt{4 x - 1}} \arccos\left( \frac{3 x -1}{2 x^{3/2}} \right),
\nn \\
R_\xi(x)  &=& -\frac{1 - x}{36 x^2} (-3 + 53 x - 541 x^2 + 407 x^3 ) + \frac{1}{6 x} (2 + 3 x + 114 x^2 - 12 x^3) \log x  \nn \\ & &+ \frac{-2 + 19 x - 80 x^2 + 156 x^3}{3 x \sqrt{4 x - 1}} \arccos\left( \frac{3 x -1}{2 x^{3/2}} \right),\nn \\
R_\Gamma(x)&=&
\frac{1-x}{72x^2}\left(3-73x+467x^2-409x^3\right)+\frac{1-15x+108x^2-12x^3}{12x}\ln
x \nn \\
&&+\frac{-1+17x-88x^2+156x^3}{6x\sqrt{4x-1}}\arccos\left(\frac{3x-1}{2x^{3/2}}\right), \nn
\end{eqnarray}
\begin{eqnarray}
R_{W}(x)  &=&  - 4 (1-x)(9 x -5) +    ( -4 + 24 x - 8 x^2) \log x     + \frac{ 8 - 64 x + 112 x^2}{\sqrt{4 x - 1}} \arccos\left( \frac{3 x -1}{2 x^{3/2}} \right), \nn \\
R_{WW}(x) &=& -\frac{4 (1 - x)}{9 x} ( 17 - 82 x + 89 x^2) + \frac{1}{3 x} (4 - 36 x + 120 x^2 -24 x^3 ) \log x  \nn \\ & & - \frac{ 8 - 88 x + 320 x^2 - 432 x^3}{ 3 x  \sqrt{4 x - 1}} \arccos\left( \frac{3 x -1}{2 x^{3/2}} \right) ,\nn \\
R_{W\tilde{W}}(x) &=& -\frac{4 (1 - x)}{9 x} ( 17 - 64 x -  x^2) + \frac{1}{3 x} ( 4 - 36 x  +24 x^2 ) \log x  \nn \\ & & - \frac{ 8 (7 x - 1)(4x -1)}{ 3 x  \sqrt{4 x - 1}} \arccos\left( \frac{3 x -1}{2 x^{3/2}} \right).
\end{eqnarray}
We do not consider Yukawa corrections to $h \rightarrow W W^*$ and $h \rightarrow Z Z^*$, since non-standard Yukawa would mainly affect the decay of a Higgs boson into jets. 
Expressions for  $\Gamma(h \rightarrow q \bar q)$  can be found in Ref.\ \cite{Spira:1997dg}.

\section{Minimal Flavor Violation}\label{app:MFV}
In this appendix we briefly discuss the Minimal Flavor Violation framework, its implications for the couplings of the operators in the SM-EFT, and its implementation into the code.

MFV is based on the flavor-symmetry group $SU(3)_Q\times SU(3)_u\times SU(3)_d$ acting on the quark fields, and $ SU(3)_L\times SU(3)_e$ acting on the leptons \cite{D'Ambrosio:2002ex}. Within the SM the above symmetry group is only broken by the Yukawa interactions. MFV makes the assumption that this remains the case when going beyond the SM, such that BSM interactions  only break the above symmetries through insertions of the Yukawa matrices which now act as spurions. Given the transformation properties of the Yukawa matrices one can then derive all possible structures for a given dimension-6 coupling. Here we take into account all  possible structures, while neglecting the small Yukawa couplings,  i.e.\ we take $y_{b,c,s,d,u}\to 0$ and $y_{\tau, \mu, e}\to 0$.

With these assumptions we find the following allowed structures for the corrections to the $W$ and $Z$ boson couplings $c_{Q\varphi, U}$, $c_{Q\varphi, D}$, $c_{D\varphi}$
and $c_{U\varphi}$
\bea
c_{D\varphi} &=& a_{D\varphi} \mathbbm{1},  \label{MFV0a}\\
\{ c_{Q\varphi, U} , c_{U\varphi}  \} &=& \left\{ a_{Q\varphi, U} , a_{U\varphi} \right\} \mathbbm{1}+ \{ b_{Q\varphi, U} , b_{U\varphi} \} \bma 0&0&0\\0&0&0\\0&0&1\ema \,,\label{MFV0b}\\
c_{Q\varphi, D} &=& a_{Q\varphi, D} \mathbbm{1}+b_{Q\varphi, D} V_{CKM}^\dagger \bma 0&0&0\\0&0&0\\0&0&1\ema V_{CKM}
.\label{MFV0c}\eea
Similarly, the allowed structures for semileptonic four-fermion operators are
\bea
\left\{ C_{ed}, C_{Ld} \right\} &=& \left\{ a_{ed}, a_{Ld} \right\} \mathbbm{1}, \label{MFV1}\\
\left\{ C_{LQ, U}, C_{eu}, C_{Lu} \right\}&=&  \left\{ a_{LQ, U}, a_{eu}, a_{Lu} \right\} \mathbbm{1}+\left\{b_{LQ, U}, b_{eu}, b_{Lu}\right\} \bma 0&0&0\\0&0&0\\0&0&1\ema,\label{MFV2}\\
\left\{ C_{LQ, D}, C_{Qe} \right\}&=& \left\{a_{LQ, D}, a_{Qe}\right\} \mathbbm{1}+ \left\{b_{LQ, D}, b_{Qe}\right\} V_{CKM}^\dagger \bma 0&0&0\\0&0&0\\0&0&1\ema V_{CKM}.\label{MFV3}\eea
The couplings
$a_i$ and $b_i$ in Eqs. \eqref{MFV0a} - \eqref{MFV3} are  free real parameters expected to be of $\Or(v^2/\Lambda^2)$, with no further Yukawa or CKM suppression. In the above equations, the structures proportional to the $b_i$ arise from insertions of the $u$-type Yukawa matrices, $\left(Y_u Y_u^\dagger\right)^n$. The $b_i$ terms in Eqs.\ \eqref{MFV0b} and \eqref{MFV2} lead to operators involving explicit top quarks, which do not have any effects in the processes considered in this paper. As a result, we do not explicitly include these pieces in the code. Note that in the MFV framework, the assumption of lepton-flavor universality, made in Section \ref{sec:flavorStructure}, simply follows from the flavor symmetries.

Finally, the couplings that do not appear in Eqs.\ \eqref{MFV0a}  -  \eqref{MFV3} can be set to zero given our assumptions. The reason is that most of these operators mix left- and right-handed quark fields, such as the $\psi^2\vp^2 X$ and $\psi^2\vp^3 $ classes as well as $C_{LedQ, LeQu}^{(1,3)}$, or couple right-handed $u$-type quarks to right-handed down-type quarks in the case of $\xi$. All of these couplings therefore come with at least one insertion of a Yukawa matrix, unlike the couplings in Eqs.\  \eqref{MFV0a} - \eqref{MFV3}. Since we neglect small Yukawa couplings this means we do not have to consider most of these interactions. The only exceptions would be the $u$-type dipole operators and the corrections to the $u$-type Yukawa couplings, for which a term proportional to the top Yukawa survives. Nevertheless, as these terms always involve explicit top quarks, we neglect them in the code.

In order to apply the MFV framework as described above the user has to set the \texttt{mfv} flag to 1. The couplings of the dimension-six operators are then constrained to the forms of Eqs.\  \eqref{MFV0a} - \eqref{MFV3}. The notation for the $a_i$ and $b_i$ coefficients that can be set in the input file are listed in Table \ref{tab:mfv}.

\begin{table}
\center
\begin{tabular}{||c|c  c|| c | c c || }
\hline
	Operator    & $a_i$ & $b_i$ & Operator & $a_i$ & $b_i$\\\hline
$	c_{Q\vp,U}$ & \texttt{A}\_\texttt{QphiU} & - &
$	c_{Q\vp,D}$ & \texttt{A}\_\texttt{QphiD} & \texttt{B}\_\texttt{QphiD}  \\
$	c_{U\vp }$  & \texttt{A}\_\texttt{Uphi} &-&
$	c_{D\vp }$  & \texttt{A}\_\texttt{Dphi}&- \\
$	C_{LQ,U}$   & \texttt{A}\_\texttt{QLu} &-&
$	C_{LQ,D}$   & \texttt{A}\_\texttt{QLd}& \texttt{B}\_\texttt{QLd} \\
$	C_{eu}$     & \texttt{A}\_\texttt{eu}&- &
$	C_{ed}$     &\texttt{A}\_\texttt{ed}&- \\
$	C_{Lu}$     &\texttt{A}\_\texttt{Lu}& -&
$	C_{Ld}$     &\texttt{A}\_\texttt{Ld}& -\\
& & & 
$	C_{Qe}$& \texttt{A}\_\texttt{Qe}& \texttt{B}\_\texttt{Qe} \\
\hline
\end{tabular}
\caption{Notation for the coefficients of the different flavor structures in the MFV framework, discussed in Appendix \ref{app:MFV}, which can be set in \texttt{POWHEG}.}\label{tab:mfv}
\end{table}

\bibliographystyle{JHEP} 
\bibliography{bibliography}

\end{document}